\documentclass{vldb}
\usepackage{graphicx}
\usepackage{url}
\usepackage{balance}  % for  \balance command ON LAST PAGE  (only there!)

\newcommand{\argmin}{\operatornamewithlimits{argmin}}

\begin{document}

% ****************** TITLE ****************************************

\title{Adaptive Cardinality Estimation}

% ****************** AUTHORS **************************************

\numberofauthors{2}

\author{
\alignauthor
Oleg Ivanov \\
       \affaddr{Postgres Professional}\\
       \affaddr{Moscow, Russia}\\
       \email{o.ivanov@postgrespro.ru}
\alignauthor
Sergey Bartunov\thanks{Author is now at DeepMind.}\\
       \affaddr{National Research University \\ Higher School of Economics}\\
       \affaddr{Moscow, Russia}\\
       \email{sbos@sbos.in}
}
\date{30 June 2016}

\maketitle

\begin{abstract}
In this paper we address cardinality estimation problem which is an important subproblem in query optimization. Query optimization is a part of every relational DBMS responsible for finding the best way of the execution for the given query. These ways are called plans. The execution time of different plans may differ by several orders, so query optimizer has a great influence on the whole DBMS performance. We consider cost-based query optimization approach as the most popular one. It was observed that cost-based optimization quality depends much on cardinality estimation quality. Cardinality of the plan node is the number of tuples returned by it.

In the paper we propose a novel cardinality estimation approach with the use of machine learning methods. The main point of the approach is using query execution statistics of the previously executed queries to improve cardinality estimations. We called this approach adaptive cardinality estimation to reflect this point. The approach is general, flexible, and easy to implement. The experimental evaluation shows that this approach significantly increases the quality of cardinality estimation, and therefore increases the DBMS performance for some queries by several times or even by several dozens of times.
\end{abstract}

\section{Introduction}

SQL is known as a declarative language. In a query user defines what operation to perform and the properties of data for this action. But the particular way for performing this operation is constructed by the DBMS. 
These ways are called query \emph{plans} in the relational DBMS. Typically the plan is represented as a rooted tree in which nodes describe physical operations with data and edges describe data flows between nodes. There may be a lot of different plans for one query and their number grows at least exponentially with the number of the joined relations. 

Since the execution time of different plans of the same query may differ by many orders, \emph{query optimization} problem arises as selection the fastest plan for a given query. Query optimizer is an important part of modern DBMS since its quality has crucial impact on the performance of DBMS.

Actually, not only execution time, but also other quality metrics can be optimized in the general query optimization problem setting. But the query execution time is the most common and natural case, so further we consider it.

The majority of modern DBMS use \emph{cost-based} approach to query optimization. This approach performs plan selection as follows: it introduces a function that estimates the \emph{cost} of the given plan and then minimizes value of this function over all possible plans. Cost is usually the amount of some type of resources (e.g., time) needed to execute the plan. The plan that minimizes the cost function is considered to be the best plan.

The first cost-based query optimizer and cost-based approach to query optimization in general were proposed in System R. So modern query optimizers based on the same principles as were proposed in System R, even if their implementation is different.

One of the main drawbacks in System R query optimization model is its \emph{cardinality} estimation (we show it below). Cardinality of the plan node is number of tuples returned by it. For cardinality estimation its is proposed to use one-dimensional histograms or indexes. Nevertheless, it is mathematically impossible to estimate cardinality somehow for correlated columns or dependent conditions using only this information. However, cardinality estimation is necessary for cost estimation and therefore for the whole cost-based optimization model. That is why it is proposed to consider clause independence in System R. Under this assumption cardinality estimation is possible and easy with given information.

The issue appears when the assumption is not fulfilled, which happens rather often. In some cases bad cardinality estimation does not cause bad plans selection, but in other cases the selected plans are worse than optimal by several orders.

In this we paper we propose a novel machine learning approach for improving cost-based query optimizers by predicting cardinalities of nodes of query execution plans. Our approach is based on the extraction of information about data and clauses dependencies from collected query execution statistics. So it adapts to the current workload and the contents of the database, that is why we call it adaptive cardinality estimation. It can be used with a variety of different machine learning algorithms from nearest neighbours to neural networks.

The paper is organized as follows.
In section~\ref{sec:cost-based} we review cost-based query optimization method which is used in the majority of modern DBMS. Also in this section we show that the part of this method which causes the most significant mistakes in choosing the fastest query plan is cardinality estimation.

In section~\ref{sec:related_work} we discuss known papers on improving cost-based query optimization model. The most popular in the DBMS community way to solve the problems with cardinality estimation is using multidimensional statistics. We discuss the drawbacks of this method. Unfortunately, the works devoted to other approaches of solving cardinality estimation problem are not provide enough experimental evaluation of their prototypes to determine whether they are adequate for industrial usage, what their drawbacks are and how to deal with them.

In section~\ref{sec:mlqo_definition} we propose our machine learning approach to cardinality estimation problem.
In subsection~\ref{sec:aqo_problem_setting} we formulate the machine learning problem for cardinality estimation.
In subsection~\ref{sec:mlqo_ml} we propose the exact machine learning method to solve the problem.
In subsection~\ref{sec:mlqo_theoretical} we discuss the theoretical properties of our approach, its possible drawbacks and possible ways to overcome these drawback in industrial usage of our approach.

In section~\ref{sec:experiments} we evaluate our approach on StrongCor, JOB, TPC--H, TPC--DS benchmarks and discuss the results.
We show three main statements in this section. First statement is that query optimizers used in modern DBMS do not choose the fastest plan, and the potentially available speed up is enormously large. Second statement is that our method allows to get nearer to optimal plans and to achieve this speed up in practice. Third statement is that our method works better for complex queries. It happens because a lot of queries with complex structure are not computationally complex but just not optimized properly.

In section~\ref{sec:future} we propose ideas of further improvement of our method and the whole query optimizer in general.

\section{Overview of cost-based query optimization}\label{sec:cost-based}

System R~\cite{Astrahan:1976:SRR:320455.320457} is known as the first relational DBMS. It's query optimization method~\cite{selinger1979access} is called \emph{cost-based} query optimization~\cite{Chaudhuri:1998:OQO:275487.275492}. It is remarkable that base principles originated from System R's query optimizer stayed unchanged in modern DBMS.

Cost-based query optimizer main principle is that for each plan we can compute somehow its cost, that is amount of resources necessary for its execution. If one can estimate cost of a plan, then finding the cheapest plan turns into the following optimization problem:
\begin{equation}\label{eq:cost_optimization_problem}
Cost(x) \to \min\limits_{x \in all\_plans(query)}
\end{equation}
The goal of algorithm is to find the best plan which joins given relations under given clauses. The algorithm used in System R for this purpose works as follows: firstly, it computes minimal cost for scanning each given relation, then it computes the minimal cost of joining for each pair of these relations, then computes the minimal cost of joining for each subset of size 3 of the given relations based on the previously computed results, and so on. Finally it computes the minimal cost of joining of all given relations. The optimal plan which corresponds to this minimal cost can be restored using backward pass, or the optimal joining plan for each subset of relations can be stored in the additional memory.

The algorithm uses the following formula of finding the cost of cheapest plan
\begin{displaymath}
R = (R1, R2, \dots, Rn)\mbox{ is a set of given relations to join,}
\end{displaymath}
\begin{displaymath}
BestCost(R) = \min\limits_{\emptyset \neq r \subset R}CostJoin(BestCost(r), BestCost(R \setminus r))
\end{displaymath}
Since best plans for joining every subset or relations is memorized and is computed only once, this algorithm is the dynamic programming algorithm. The worst memory complexity of algorithm is $O(2^nn)$, the worst time complexity is $O(3^n)$. Remarkable part of this algorithm is its guarantee of finding the cheapest plan.

Nowadays SQL--queries may contain much more than a dozen of relations to join. That is why modern DBMS also use approximate methods for solving problem~\ref{eq:cost_optimization_problem}. As an example of such methods we can mention genetic algorithm~\cite{Mitchell:1998:IGA:522098} and simulating annealing~\cite{kirkpatrick1983optimization}.

Nevertheless the majority of queries has not a lot of relations to join. This fact and guarantees of optimality for dynamic programming as optimization method draws the conclusion, that cost estimation has the most influence on the query optimization.

The cost models are used for the estimation of the cost of the plan node. For example, if we have in-memory \texttt{sort} node we estimate its cost as $1.39 \cdot n\log_2 n \cdot c_o$, where $1.39$ is an average constant for comparisons in quick sort algorithm, $c_o$ is cost of a CPU comparison operation and $n$ is a number of tuples to be proceeded in this node. In a similar way one can construct cost models for all kind of nodes.

Number $n$ of node's returned tuples is called cardinality of the node. It is the only part in cost model, which depends not on hardware parameters but on data in database and clauses in the query.

We performed a research to determine which part leads to the most significant errors: cost models (formulas and hardware constants) or cardinality estimation. You can see the details in appendix~\ref{sec:cardinality_or_cost}. We conclude in that research that cardinality estimation is much more critical than the cost model, especially in the case of complex queries.
Independently the same results were obtained in the recent work~\cite{Leis:2015:GQO:2850583.2850594}.

%возможно следует выделить подраздел для описания selectivity и cardinality (not fixed)

Hence we focus on cardinality estimation problem. For cardinality estimation DBMS must combine the information about the clauses and the data in the database. Histograms are an easy and a fast way to store the information about the data distribution. In the majority of DBMS histograms are built on single columns. Using histograms one can easily and precisely enough estimate the number of tuples in table which fulfills the range clause or equal clause.

Clause selectivity in the node is the ratio of tuples in the node which fulfills the clause to the total number of tuples proceeded in the node. We also may consider selectivity of clause in node as probability of a tuple in node to fulfill the clause.

The node selectivity is the ratio of tuples in the node which fulfills all clauses in it to the total number of tuples proceeded in the node. We also can say that the node selectivity is the probability of a tuple in node not to be filtered. If we know the selectivity of the node and the cardinalities of its children for the join node or the number of tuples in the table for the scan node, we can compute cardinality of that node.

The problem is how to compute the node selectivity. Using one-dimensional histograms one can estimate only clause selectivities. In probability theory terms, we know marginal probabilities of a tuple to fulfill each clause, but want to estimate the probability of a tuple to fulfill all clauses simultaneously which is an ill-posed problem.

Consider an example where in the table with schema \texttt{(a int, b int)} there are 10000 tuples (0, 0) and 10000 tuples (1, 1). For this kind of data selectivity of clauses \texttt{a = 0} and \texttt{b = 0} is equal to 0.5 and the selectivity of node with both these clauses is also 0.5.
Now consider the table with 10000 tuples (0, 1) and 10000 tuples (1, 0). In this case the histograms and selectivities of clauses are the same, but the node selectivity becomes 0, because for this table clauses \texttt{a = 0} and \texttt{b = 0} are mutually exclusive.
So clause selectivities don't contain enough information for the node selectivity estimation.

To solve the specified ill-posed problem the assumption of clauses independence is accepted in the majority of modern query optimizers. The clauses independence assumption means that trueness of one clause does not depend on the trueness of all other clauses. This implies that the node selectivity is equal to the product of selectivities of all clauses in that node.

Obviously, on real databases the independence assumption often fails. That may lead to both underestimation and overestimation of nodes cardinality. The most common case is underestimation, because in real SQL-queries the clauses more often are positive-correlated than negative-correlated.

Thus, we consider the problem of correct node cardinality estimation as the best way to improve cost-based query optimizer.

\section{Related work}\label{sec:related_work}

There is no perfect solution for the query optimization problem, so a lot of paper on existing algorithms improvement are published. There are different lines of research in the query optimization field.

One line of research is devoted to the query execution time prediction using machine learning methods: \cite{Ganapathi:2009:PMM:1546683.1547490}, \cite{6228100}, \cite{Hasan:2014:MLA:2682647.2682729}, \cite{Li:2012:RER:2350229.2350269}, \cite{malik2007black}. In these works the predictions are made only for those plans, which have been chosen by the query optimizer. Therefore, they applicable only for the time prediction on the subset of all plans in which every plan may be chosen by query optimizer:
\begin{displaymath}
\big\{\argmin\limits_{x \in all\_plans(query)}Cost(x) \;\big|\; query \in Queries\big\}
\end{displaymath}
Nevertheless, for the query optimization we have to predict time not only for those plans which may be chosen for execution. That is why these works are non-applicable for improving the query optimizer.

The second line of papers~\cite{Wu:2013:TPQ:2536206.2536219}, \cite{6544899}, \cite{DBLP:journals/corr/WuWHN14}, \cite{DBLP:journals/corr/WuNS16} use sampling-based approach.
Like the previous group, they do not improve query optimizer's cost estimator directly, but propose sampling-based ways to rectify standard query optimizer's errors if they are. Sampling-based approaches are good on low-relational queries, but on queries with lots of joins they may have too large variance.

In the third line of papers authors try to improve the query optimizer by improving its cost model.
In~\cite{6544899} least squares method is used for tuning PostgreSQL cost model parameters. In~\cite{6228100} authors propose a linear cost model with a large amount of features. We believe that existing cost models are not perfect, but improving cost model is not affect DBMS performance as much as improving cardinalities, so we do not consider these works here.

The fourth line of works is about using multidimensional statistics for correct cardinality estimation: ~\cite{Gunopulos05selectivityestimators}, \cite{poosala1997selectivity}, \cite{poosala1996improved}, \cite{Bruno01stholes:a}, \cite{Furtado99summarygrids:}, \cite{yu2006hase}, \cite{Halim:2009:FEH:1645953.1646101}. The most popular kind of such statistics are multidimensional histograms. It is the most popular way to improve standard cardinality estimation method in DBMS community.

The main drawback of multidimensional histograms is their memory complexity and time complexity of building and accessing such histogram. This complexity grows exponentially of the number of dimensions, so only low-dimensional histograms are available. Therefore we face the second drawback with these histograms: we have to determine on which columns set to build them. This problem requires good database administrator or complicated software tool which analyses queries and use statistical tests to check the correlation between columns.

Also in~\cite{poosala1997selectivity} authors propose using singular value decomposition for selectivity estimation for range clauses, but this approach works only with real number and is not very flexible. That is why it has not been implemented in any DBMS to our best knowledge. \cite{yu2006hase} combines sampling-based approach and multidimensional histograms.

Finally the fifth line of research use machine learning methods for cardinality estimation. There are a few papers in this line. They are described below.

In~\cite{Lakshmi:1998:SEE:645924.671200} neural networks are used to estimate the selectivity for user defined functions and data types. Nevertheless, this paper address the problem of estimation the selectivity for a single clause. That is why it is not devoted to our research.

In~\cite{Getoor:2001:SEU:376284.375727} it is proposed to estimate node selectivity using inference in the automatically constructed Bayesian networks. Authors fit Bayesian network to the data from the database using maximum likelihood principle. For building network structure various heuristics are used. The significant drawback of this method is its lack of scalability. Authors mention in the paper, that method is unusable with large amounts of data. This makes it unsuitable with modern databases. Also the problem of automatic construction of Bayesian networks on given data is not considered to be solved, that is why the quality and stability of authors solution is not clear. Also there are no experiments on incorporation of the proposed selectivity estimation method into query optimizer in the paper.

The paper~\cite{heimel2009bayesian} is applying Bayes' formula for node selectivity estimation using multidimensional statistics on the data. This paper contains a lot of proposals and heuristics which are no by means always works on real data. This paper does not address queries with joins. Also there are no experimental evaluation of this method for the query optimization in the paper.

The paper~\cite{Liu:2015:CEU:2886444.2886453} proposes to use neural networks for joint selectivity estimation of several range clauses. In this paper neural networks are similar to multidimensional histograms, but neural networks are not under the dimensionality curse. Queries with joins are not considered in this paper. Unfortunately, there are not enough information in this paper about how much time network training takes. Obviously, with long enough training on fixed data and big enough network one can obtain nearly perfect selectivity predictions. Nevertheless practical value of this method depends on how much time and how many queries are needed to obtain precise enough neural network. Unfortunately, there are no answers on these questions in the paper. Also the exact network architecture (including number of units on hidden layer) is not provided in the paper.

\section{Adaptive cardinality estimation}\label{sec:mlqo_definition}

\subsection{Machine learning problem setting}\label{sec:aqo_problem_setting}

The main principle of adaptive query optimization is using query execution statistics from the previously executed queries. So in general the problem formulated as follows: to predict cardinality for a plan node having plans for the previously executed queries and true cardinalities for their nodes.

The challenge in this problem is that plan may be an arbitrary complex binary tree with arbitrary complex clauses in its nodes. This is rather complicated structure, and no machine learning techniques can work straightforward with such objects. In this section we propose the way to overcome this limitation.

Machine learning tries to find regularities in data. The majority of machine learning methods works under the following assumptions: the given data is a set of \emph{objects} and their true \emph{target values}, respectively. Objects are represented as real-value vectors from $\mathbb{R}^n$ for some $n$. These vector are called \emph{feature vectors} and the space $\mathbb{R}^n$ is called the \emph{feature space}. Target values may belong to different spaces. The most common cases are in which target value belongs to a set of discrete numbers $\{1, 2, \dots, M\}$ for classification problem and in which target value is a single real number from $\mathbb{R}$ for regression problem. We want to predict cardinalities of plans nodes, so for our problem the second case is more suitable. The machine learning method uses given data to build an algorithm which can predict target value for a new object.

We defined our target value, it is the cardinality of plan node. To completely reduce cardinality estimation problem to machine learning problem we have to introduce the feature space. To predict plan node cardinality we have to know what relations it joins and under what clauses. On one hand, in general we don't know clauses semantic and structure, so we can only use them as an atomic objects (i. e. we can only check whether two clauses are equal). On the other hand, the number of different clauses is too big even to store them all, not to process. That is why we propose the following solution: we consider that the clauses are equal if they differ at most in their constants. For example we consider that the clauses \texttt{age < 25} and \texttt{age < 26} are equal, because they both have structure \texttt{age < \textbf{const}}. This approach significantly reduces number of different clauses, but has an obvious drawback: clause constants contains a lot of information about cardinality. To take into account the information about constants, we compute the selectivities of each clause using standard DBMS method with histograms. The vector of selectivities is the feature vector for in out problem setting. The only exception is the classes of equivalence with size larger than two on variables: in this case pairwise clauses may be different for the same restrictions, so we consider it as one set equality operator without selectivity.

The point of previous transformation is to map nodes into the real space, because the majority of machine learning methods works in the finite real feature space.

We haven't developed techniques to extract semantic features from clauses, classes of equivalence or base relations. That is why we consider that they are not comparable and for each set of constant-deprivated clauses, classes of equivalence and base relations we introduce its own feature space. That solutions has pros and contras. Pro is the high stability of such method. Contra is potentially big number of such feature spaces. Theoretically we can limit number of feature spaces with the total number of executed nodes. In practice for queries with statical structure this limitation is significantly larger that the real number of obtained feature spaces.

If our method is asked for the prediction in the feature space in which we have no previously executed plans as points, it deny to make a prediction and return the cardinality from the standard DBMS estimator. This situation happens rather often during the query optimization, because to execute one plan the query optimizer has to consider much amount of other plans, the majority of which has feature space with no plans executed.

\begin{figure}
\centering
\includegraphics[width=\linewidth]{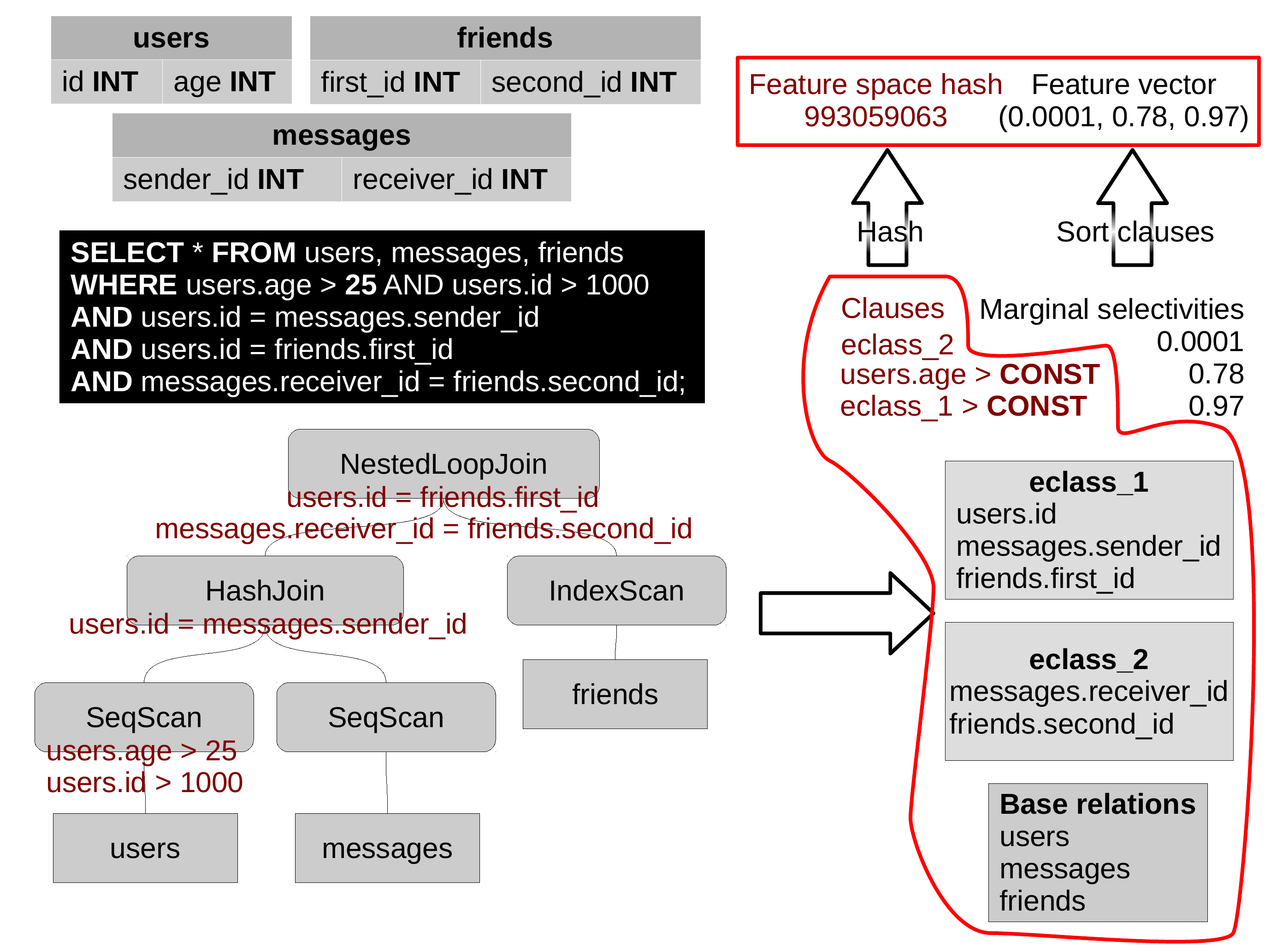}
\caption{Building machine learning feature space}
\label{fig:ml_fspace}
\end{figure}

\begin{figure}
\centering
\includegraphics[width=\linewidth]{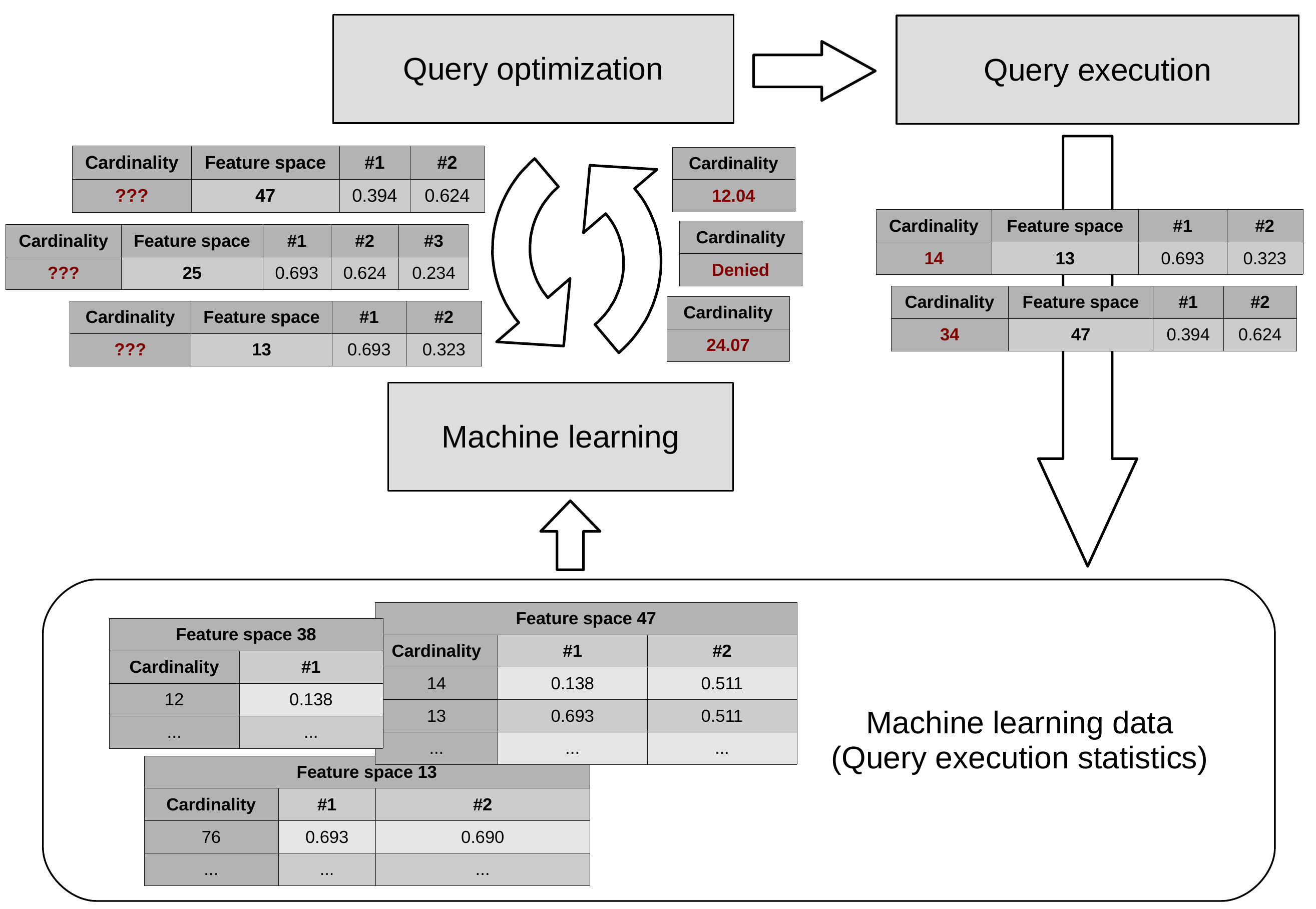}
\caption{Adaptive cardinality estimation workload}
\label{fig:ml_workload}
\end{figure}

The feature space introduction is illustrated on figure~\ref{fig:ml_fspace}. The whole process of adaptive cardinality estimation is illustrated on figure~\ref{fig:ml_workload}.

\subsection{Machine learning methods}\label{sec:mlqo_ml}

The fact is that data distribution in a feature space depends on data in database and queries. That means that the distribution may be arbitrary and arbitrary complex. That is why we have to use method that can potentially approximate the distribution of any complexity. For example, ridge or linear regression~\cite{hoerl1970ridge} is not reach enough class of regressors.

Another desirable property of the estimator is the possibility of online learning, i. e. the possibility to learn new data fast, without completely rebuilding the estimator and without saving all training data.

Also it is better for machine learning methods to take logarithm of all values. So further in this paper we predict logarithm of cardinality instead of cardinality itself, based on logarithms of selectivities of clauses.

In subsection~\ref{sec:experiments_strongcor} we performed proof-of-the-concept test of different (not necessarily online) estimators. Based on its results we decided to use use our regression modification of k nearest neighbours regressor which is described in appendix~\ref{sec:OkNN}.

\subsection{Theoretical properties}\label{sec:mlqo_theoretical}

We proved some theorems to get an insight about how the proposed method has to work.

For our theorems we assume \emph{static workload}. Static workload is workload where data does not change and number of different queries is limited. Also we assume that for each point in our constructed feature subspace there are the only one correct answer, i. e. the problem is not ill-defined after our transformation. And the third assumption is that optimization method is such that for plan choice the optimization method requests cardinality estimation for each node of that plan and uses obtained results without any changes.
% need more restrictions on the optimization method

We consider a process on each step of which one query is executed. For query optimization on each step we use the proposed method. After each step we update execution statistics. Also we suppose that number of executions is not limited for each query.

Under these assumptions we have the following conclusions about the considered process:
\begin{itemize}
  \item Plans for all queries will converge in finite number of steps, i. e. their plans will stop changing.
  \item If the machine learning method is reach enough (i. e. it can approximate arbitrary complex distribution), then cardinality predictions for the plan to which the method converges will be nearly perfect.
  \item With the perfect cost model it is guaranteed non-deceleration of query execution performance after convergence with the proposed method.
  \item To find the fastest plan possible in the common case we have to use plans space exploration techniques.
\end{itemize}

\section{Experiments}\label{sec:experiments}

\subsection{Experimental setup}\label{sec:experimental_setup}

In this paper we use PostgreSQL (version 9.6) as an example of relational DBMS with standard cost-based query optimizer.\footnote{The source code of the proposed method implementation is available at \url{https://github.com/tigvarts/aqo}.} Nevertheless our method is general and can be easily implemented in another DBMS.

We disabled parallel query execution to simplify the implementation of query execution statistics collection.
Also we figured out that PostgreSQL cost model works bad with cardinalities less than 1, so we return maximum of 1 and predicted cardinality.
To compute planning and execution time we insert \texttt{EXPLAIN ANALYSE} before each query.

The computational overhead is yet to be estimated, but it is not significant.

\subsection{Benchmarks}

For testing purposes we used 4 different benchmarks: StrongCor~\cite{Liu:2015:CEU:2886444.2886453}, TPC--H~\cite{tpch}, TPC--DS~\cite{tpcds} and Join Order Benchmark~\cite{Leis:2015:GQO:2850583.2850594}.

StrongCor is the benchmark for evaluation of the proposed cardinality estimation algorithms. This benchmark contains the database generation algorithm and the algorithm for queries generation. The benchmark is not suitable for testing DBMS performance because its queries contain no joins.

TPC--H and TPC--DS benchmarks are well-known DBMS and hardware performance benchmarks. These benchmarks contains the database generation algorithm and the algorithm for generation queries with a number of different pre-defined structures. The algorithms may be used with the different scale factors for the database size. We use both benchmarks with the scale factor 1Gb for the testing convenience. The queries in these benchmarks are known as complex analytic queries, so they are challenging for the query optimizer.

TPC--DS qualification benchmark is TPC--DS benchmark with scale factor 1Gb. The results with this scale factor may not be considered as TPC--DS benchmark results. Nevertheless our goal in this paper is not to make an appropriate TPC performance test of PostgreSQL, but to investigate the possibilities of DBMS speed up with the use of the machine learning methods. For this purposes TPC--DS qualification benchmark is suitable, but please note that obtained results are incomparable with other TPC--DS results. For the notational convenience we will call TPC--DS qualification benchmark as TPC--DS, but that is not quite correct.

Join Order Benchmark was proposed in~\cite{Leis:2015:GQO:2850583.2850594} with the detailed exploration of the query optimization bottlenecks. This benchmark contains the snapshot of IMDB database and a number of static queries to it with the different numbers of joins.

\subsection{Choosing machine learning method}\label{sec:experiments_strongcor}

In this subsection we compare the applicability of different machine learning techniques to the cardinality estimation problem. For doing that we use StrongCor benchmark which was introduced in~\cite{Liu:2015:CEU:2886444.2886453}. This benchmark in not a DBMS performance test: each its query contains the only scan of a single table which performance does not depends much on the cardinality estimation. Nevertheless this benchmark is suitable for evaluation of different machine learning methods and can be performed even without any DBMS.

We choose six machine learning methods for the comparison: linear regression~\cite{hoerl1970ridge}, neural networks~\cite{bishop95neural} and gradient boosting over decision trees~\cite{friedman2001greedy}, $k$ nearest neighbours~\cite{bishop2006patternkNN}, $k$ nearest neighbours with storing only recent $K$ objects, and our modification of $k$ nearest neighbour regression decribed in appendix~\ref{sec:OkNN}.

We choose the best parameters for each method based on balance of quality and performance. For linear regression we use stochastic gradient descent learning procedure with 10 iterations on one object. For neural network we use sigmoid activation function, 50 neurons on the hidden layer, 256 objects in the batch, and RMSProp~\cite{tieleman2012lecture} method for learning. For gradient boosting over decision trees we use maximal tree depths equal to 8.  For k nearest neighbours we use $k = 3$. For $k$ nearest neighbours with limited number of stored objects $K$ we use $K = 500$. %For lat two methods we also used objects selection heuristics: these methods has to store training set. For memory and training time optimization we do not store those objects, on which the prediction of model is already good.
%The details of choosing the best parameters for each method are in appendix~\ref{sec:ml_tuning}.
\begin{figure}
\centering
\includegraphics[width=\linewidth]{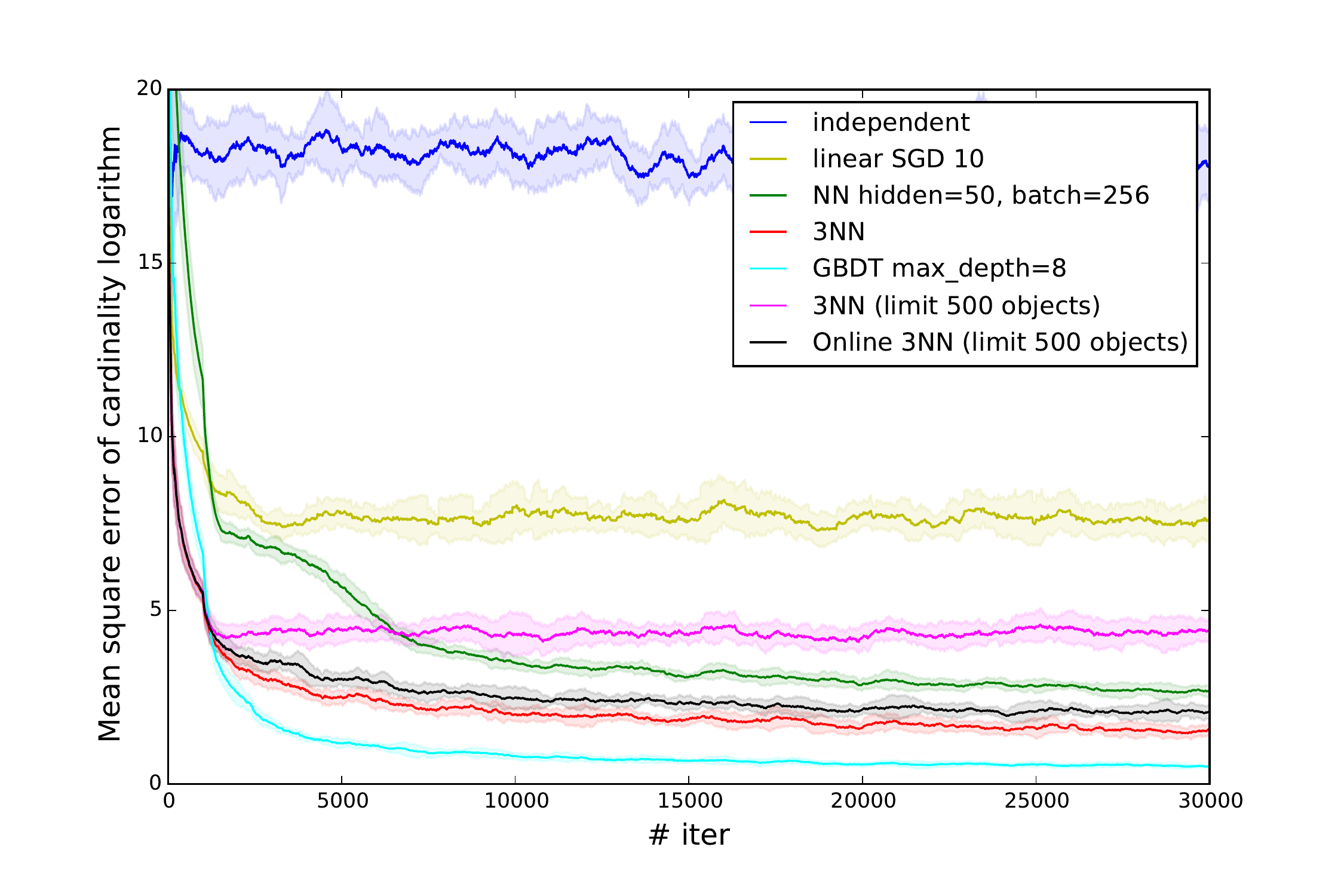}
\caption{Quality of cardinality estimation on different methods over time}
\label{fig:compare_01}
\end{figure}
\begin{figure}
\centering
\includegraphics[width=\linewidth]{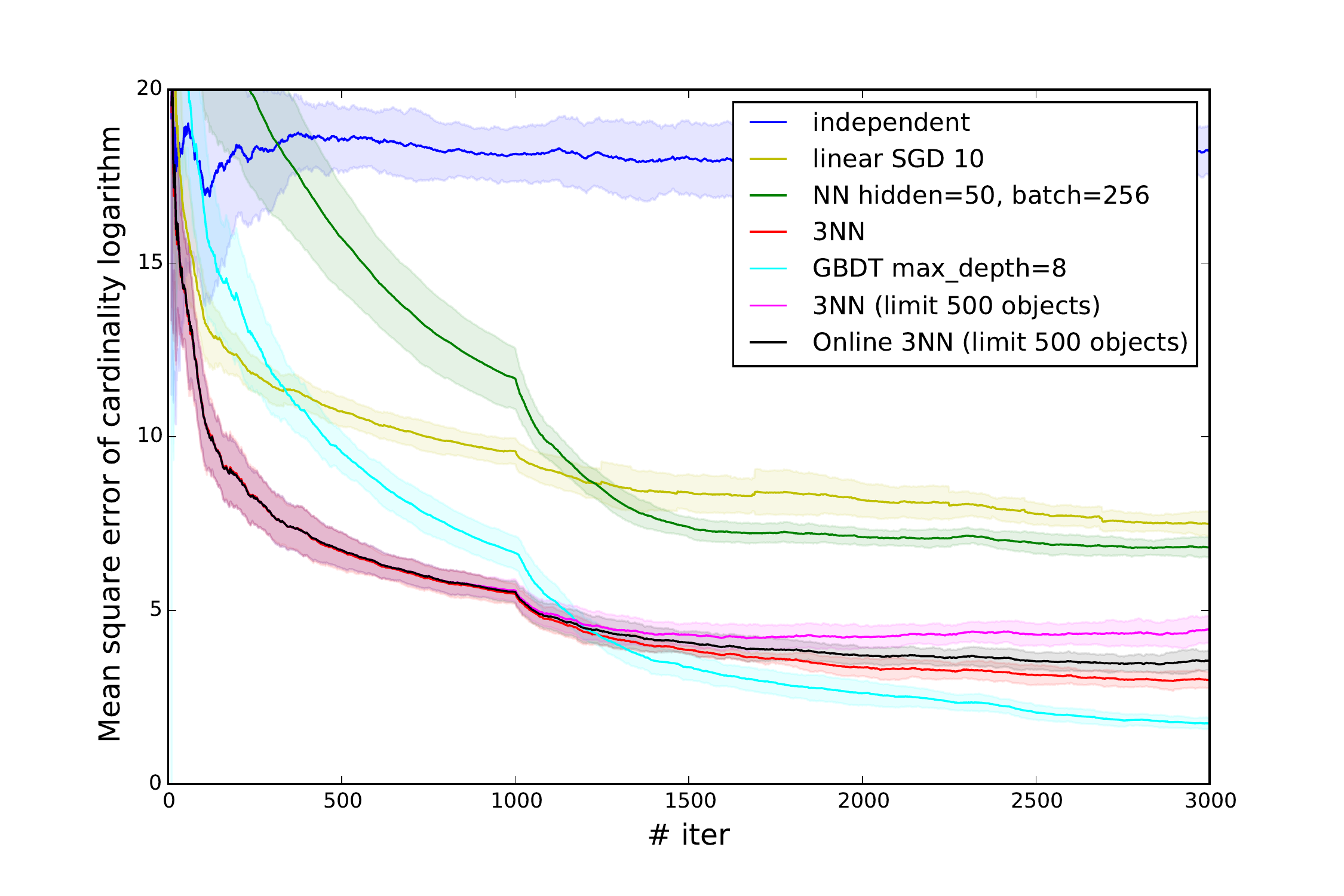}
\caption{Quality of cardinality estimation on different methods}
\label{fig:compare_02}
\end{figure}

On figure~\ref{fig:compare_01} we can see that every machine learning methods works better than standard independence assumption. Also linear regression works significantly worse than other methods, because the it can approximate only a limited family of distributions, while other methods theoretically can approximate the distributions of an arbitrary complexity.

On figure~\ref{fig:compare_02} we can see that $k$ nearest neighbours works significantly better with small amounts of data. This property is important for us, because we usually retrieve little information for slow and complex queries. Nevertheless, $k$ nearest neighbours requires to store all training objects, which requires too much memory, so it cannot be applied straightforwardly. That is why we compare following two methods: $k$ nearest neighbours which stores only last $K$ recent objects and our modification of $k$ nearest neighbours~\ref{sec:OkNN}. Both of these methods works similar to standard $k$ nearest neighbours method while number of training set objects is less than $K$. On figure~\ref{fig:compare_01} we see that our modification works better after $K$ training objects are obtained. That is why we use our modification of $k$ nearest neighbours for experiments below.

Nevertheless, one can easily use his favorite machine learning technique instead of our method in the proposed approach to predict the cardinality.

\subsection{Performance evaluation}\label{sec:experiment_benchmarks}

In this section we provide results for performance benchmarks, i. e. TPC--H (22 types if queries) and TPC--DS (99 types of queries, but 4 of them didn't run on PostgreSQL, so 95 left).

As it is stated in this section, the imperfection of cost model causes performance decrease sometimes. For simple queries the computational overhead also may avoid the benefits of our method. That is why users wish to enable or disable our method for different types of queries. We consider that two queries belongs to the same type if they differ at most in their constants. For this reason we investigate the behaviour of our method for each type of queries separately.

We divided types of queries into parts according to their execution time. Very fast are queries that execute less that 1 second, fast are queries with execution time between 1 and 10 seconds, normal are queries with execution time from 10 to 100 seconds, slow queries are with execution time from 100 seconds to 1000 seconds, the slower queries are considered to be very slow.

\begin{figure}
\centering
\includegraphics[width=\linewidth]{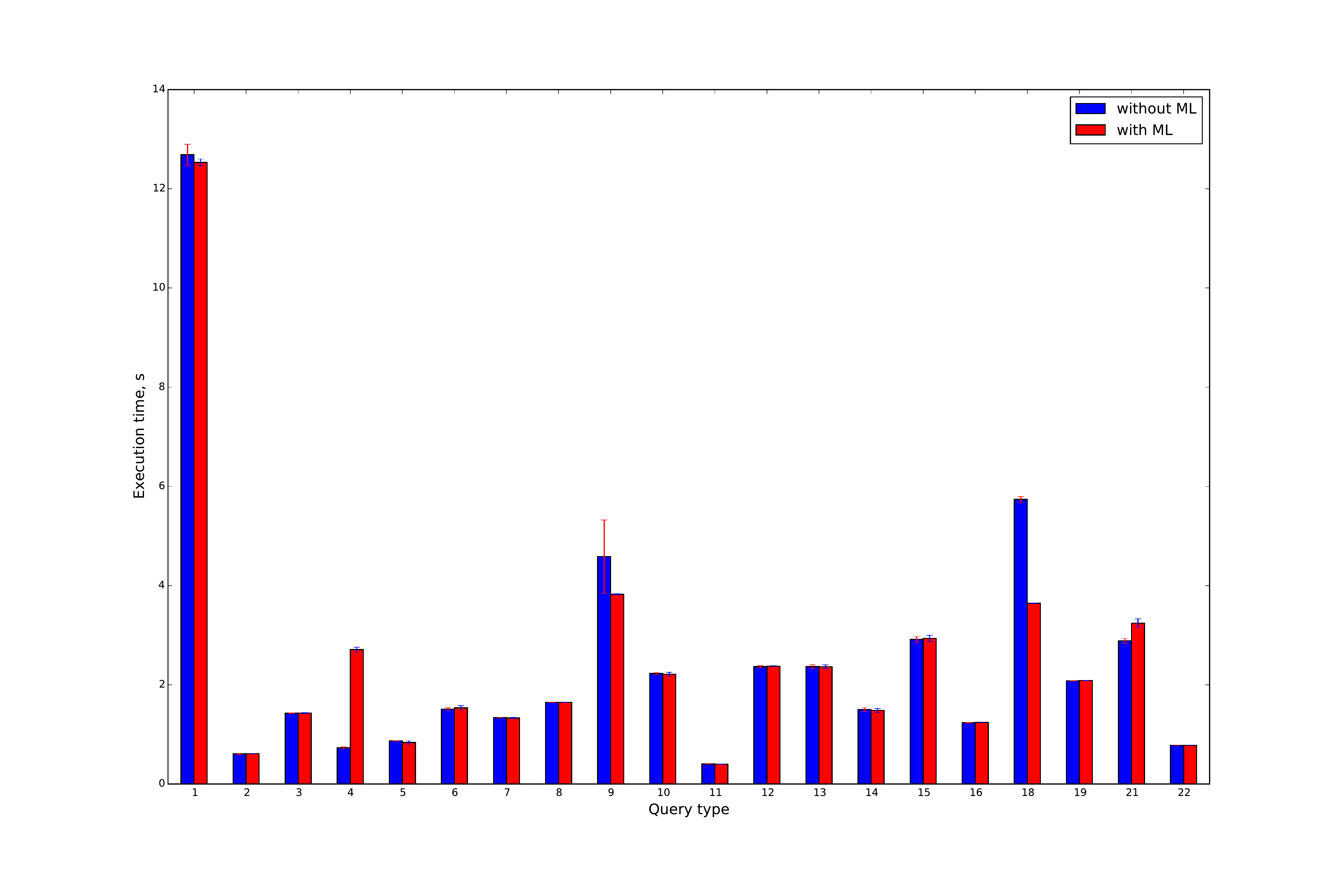}\\
\includegraphics[width=\linewidth]{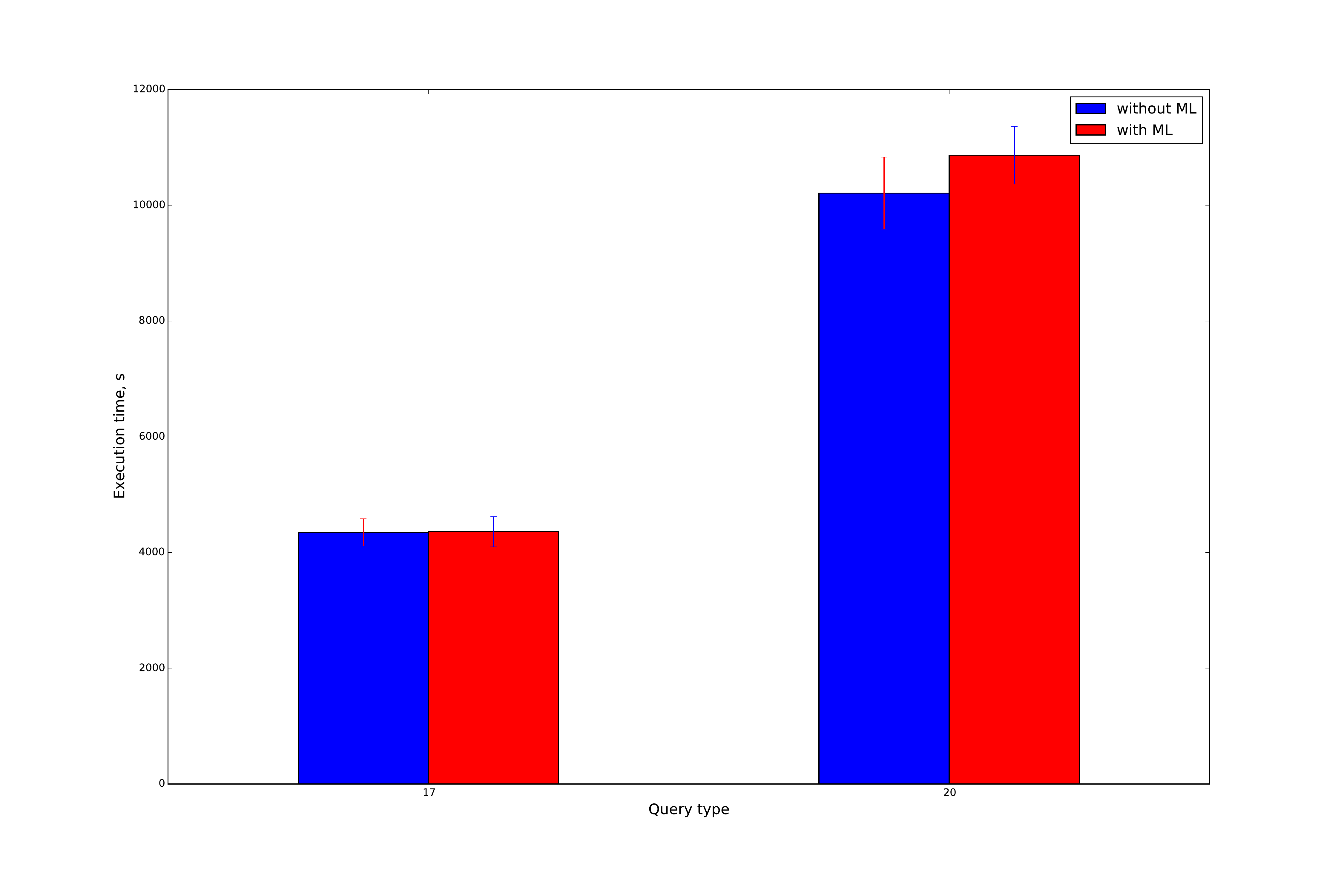}\\
\caption{Query execution time on TPC--H before and after using adaptive cardinality estimation}
\label{fig:tpch}
\end{figure}

Plots for results on TPC--H with scale factor 1Gb are provided on figure~\ref{fig:tpch}. We can see, that for 18-th query type the performance was increased, but for 4-th query type performance was decreased which is the fault of cost model.

\begin{figure*}
\centering
\includegraphics[width=0.49\linewidth]{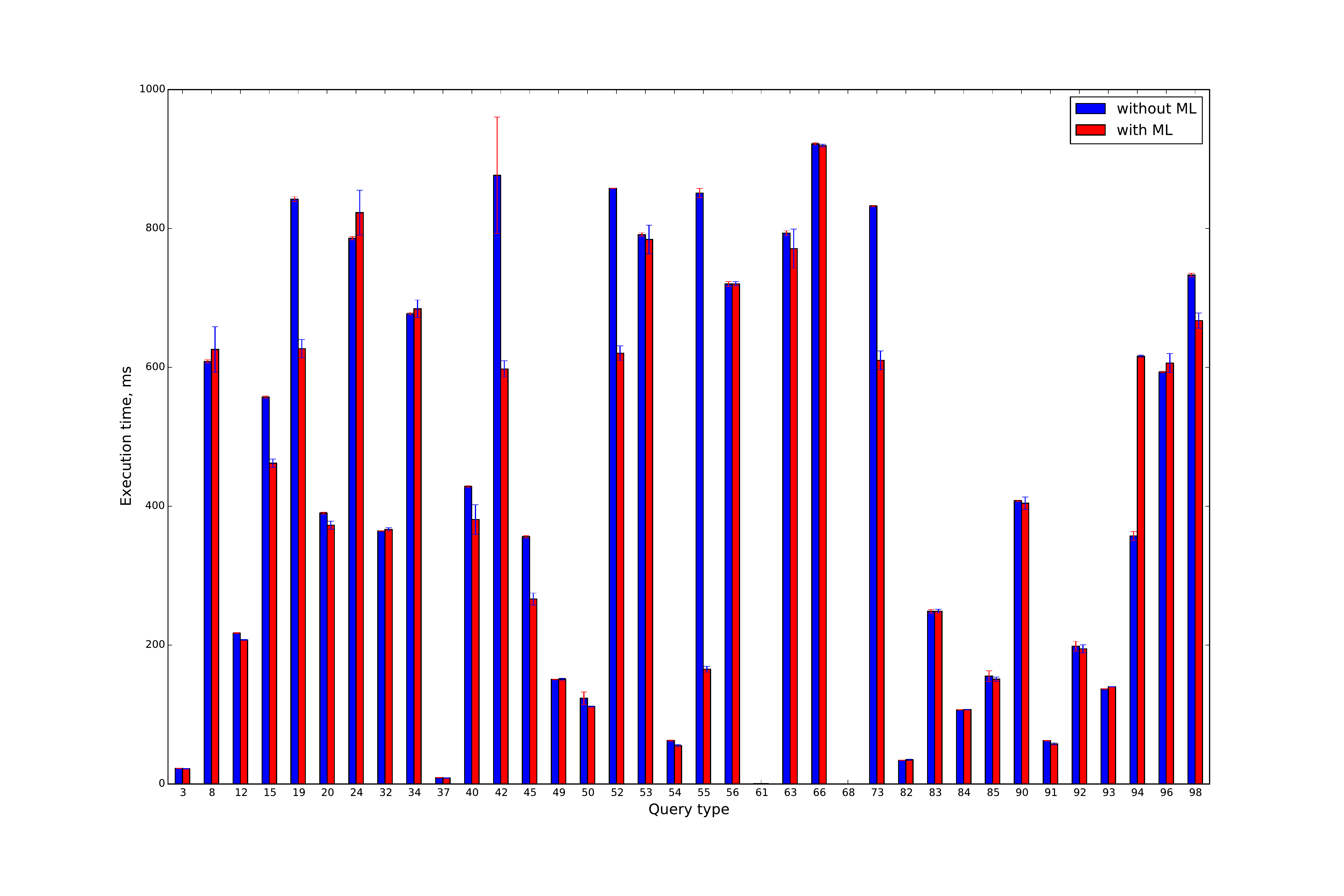}\hfill%\
\includegraphics[width=0.49\linewidth]{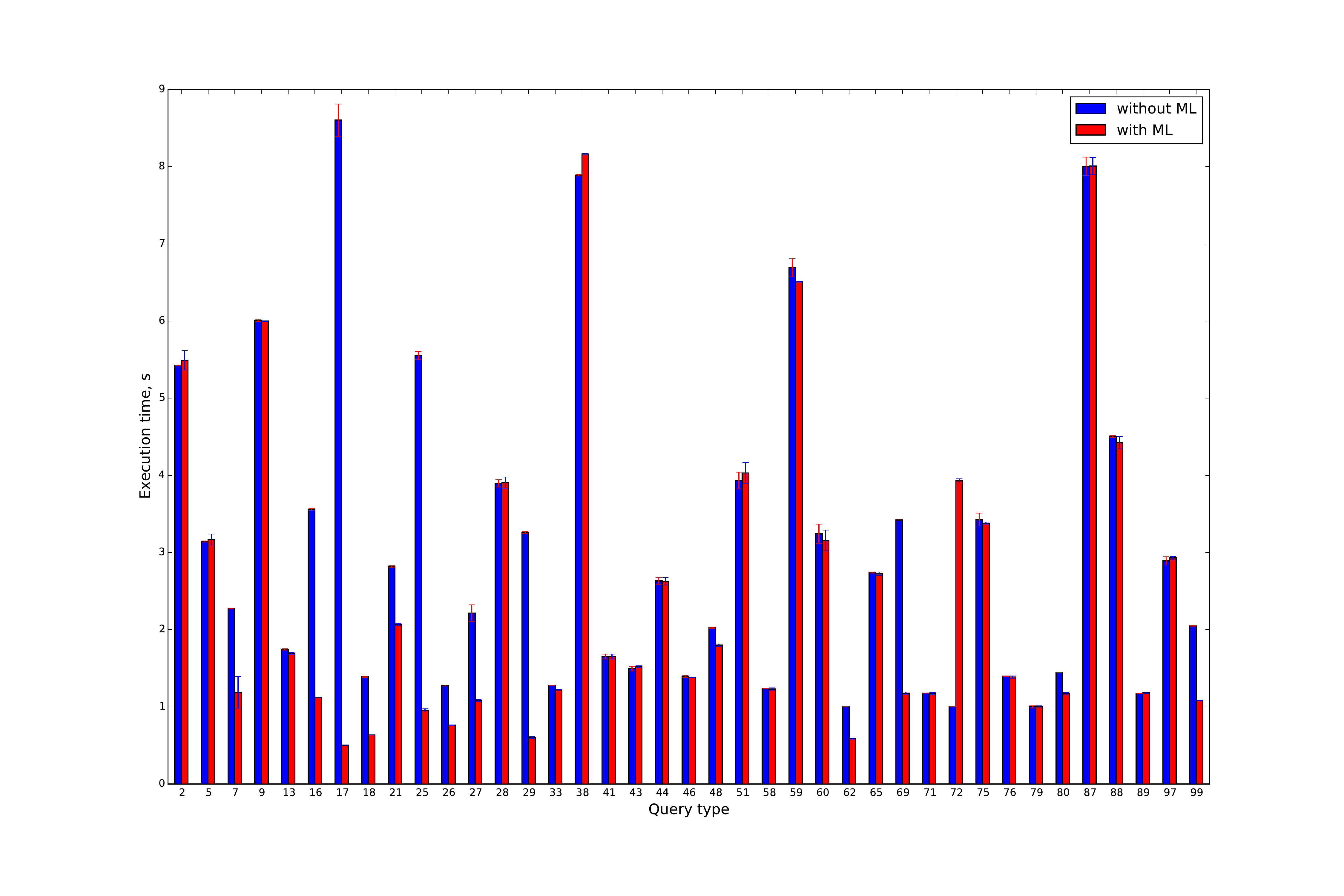}\\
\includegraphics[width=0.49\linewidth]{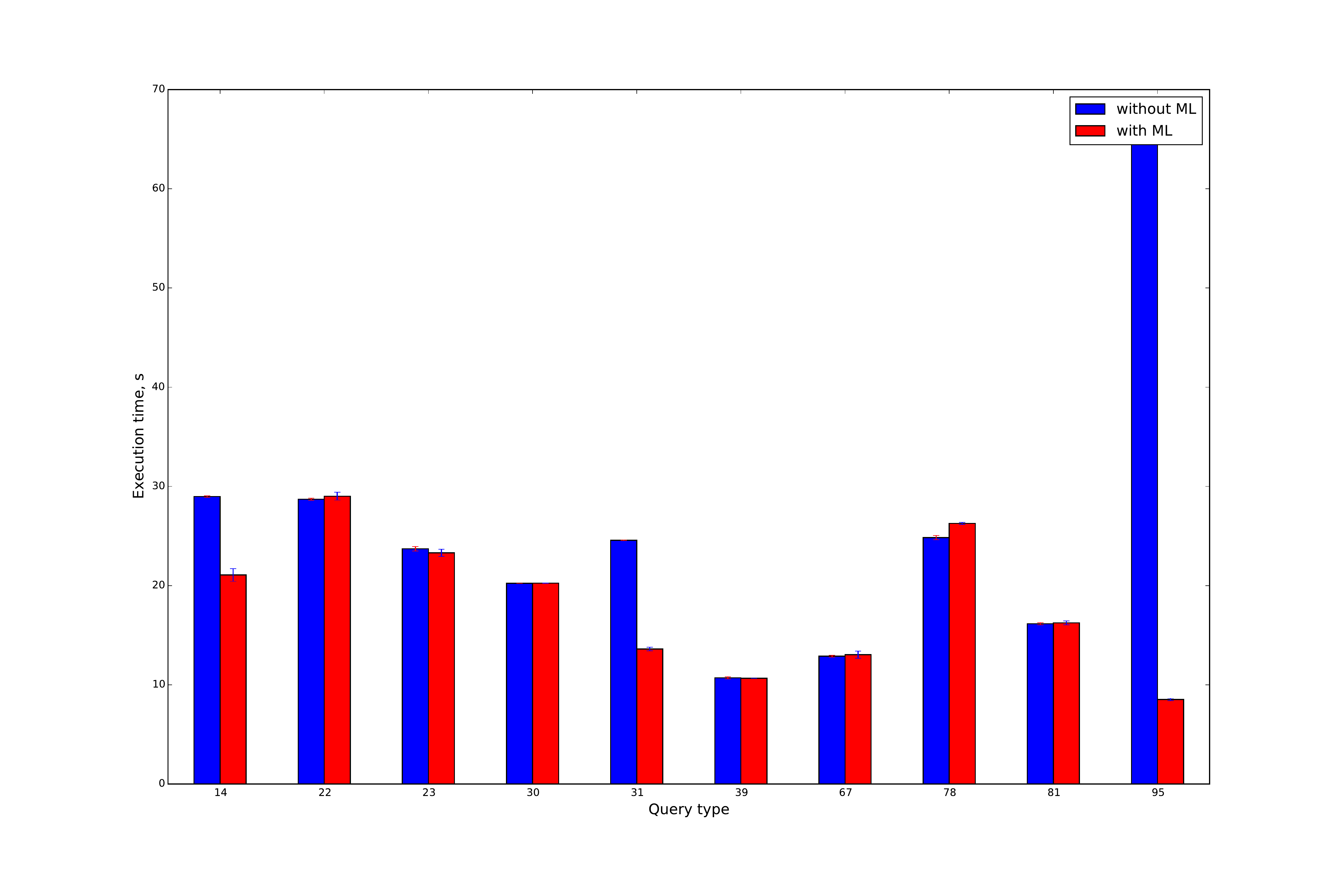}\hfill%\
\includegraphics[width=0.49\linewidth]{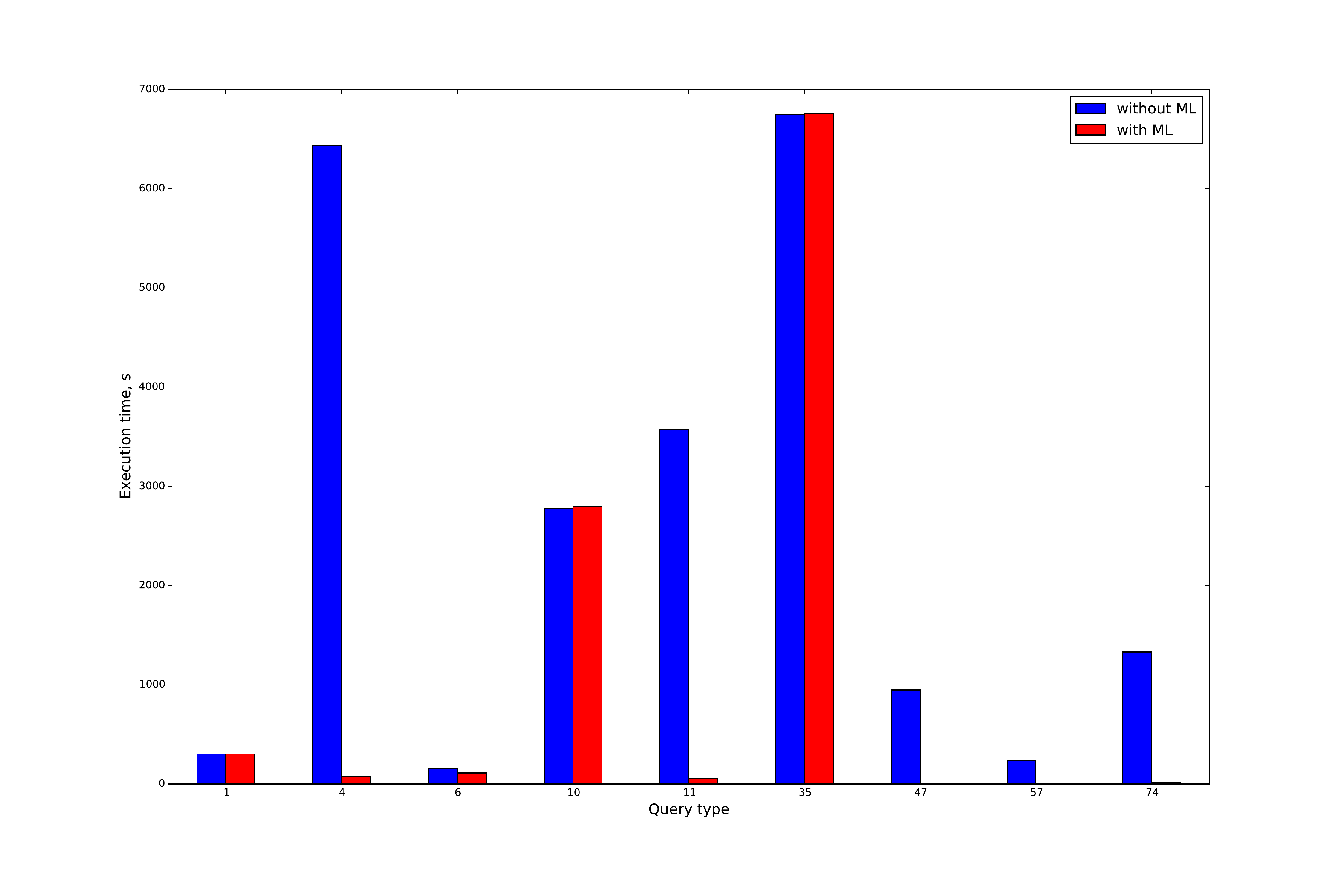}\\
\caption{Query execution time on TPC--DS before and after using adaptive cardinality estimation}
\label{fig:tpcds}
\end{figure*}

Plots for results on TPC--DS with scale factor 1Gb are provided on figure~\ref{fig:tpcds}.

\begin{figure*}
\centering
\includegraphics[width=\linewidth]{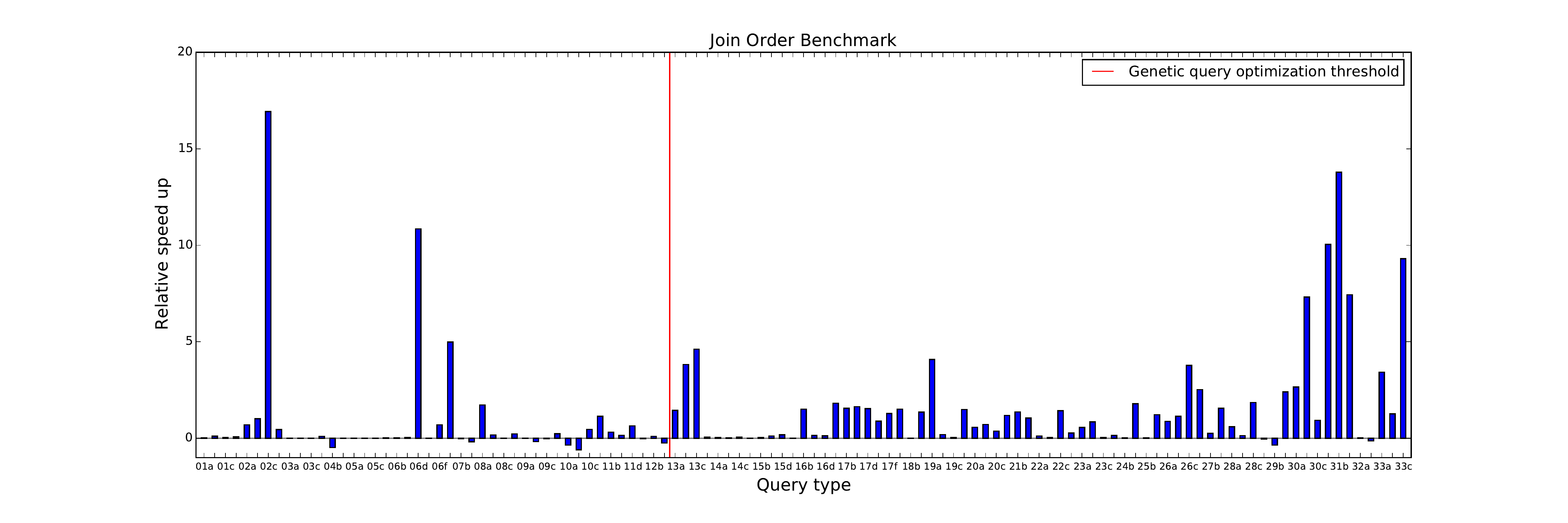}
\caption{Relative query execution time decrease after using adaptive cardinality estimation}
\label{fig:job}
\end{figure*}

Speed ups for Join Order Benchmark are plotted on figure~\ref{fig:job}.

\begin{table*}
    \centering
    \caption{The statistics on performance benchmarks}
    \label{tab:expr_res}
    \begin{tabular}{c|c|c|c|c}
        \textbf{Query group} & \textbf{Time without} & \textbf{Time with} & \textbf{Increase of} & \textbf{Maximal execution} \\
         & \textbf{proposed method, s} & \textbf{proposed method, s} & \textbf{average performance} & \textbf{time decrease} \\
                \hline
        TPC--H fast & 49.90 & 49.22 & 1.3\% & 1.57 \\
        TPC--H slow & 14559 & 15227 & -4.4\% & 1.00 \\
        TPC--DS very fast & 15.275 & 13.582 & 12.5\% & 5.15 \\
        TPC--DS fast & 119.89 & 96.64 & 24\% & 17.08 \\
        TPC--DS normal & 257.4 & 182.0 & 41.4\% & 7.83 \\
        TPC--DS slow & 1649 & 428 & 285\% & 96.69 \\
        TPC--DS very slow & 20862 & 9709 & 115\% & 95.25
    \end{tabular}
\end{table*}

In the theoretical section we mention that with perfect cost model we can expect execution time to not-increase at the end of the learning with our method. Unfortunately, PostgreSQL cost model is not perfect, so sometimes execution time grows, but in the majority of cases this increase is not significant because PostgreSQL cost model is good enough. Nevertheless, much more often case with our method is significant execution time decrease.

The aggregated statistics is available in table~\ref{tab:expr_res}. We can make the following conclusions from these results:
\begin{itemize}
  \item There are cases where our method increases quality of cardinality estimation, but decreases DBMS performance. That happens because of not precise enough cost model. We can see, that cardinality prediction quality was increased or not-decreased for all queries in benchmarks.
  \item For an average, the method increases DBMS performance.
  \item This increment is more significant for slow and complex queries. It turns out that some complicated queries are not really that complicated but just under-optimized.
\end{itemize}

\subsection{Learning dynamics}\label{sec:experiment_dynamics}

In previous section we compared the performance without our method and performance after our method is completely learned. In this section we observe the DBMS performance during learning of our method.

\begin{figure*}
\centering
\includegraphics[width=0.3\linewidth]{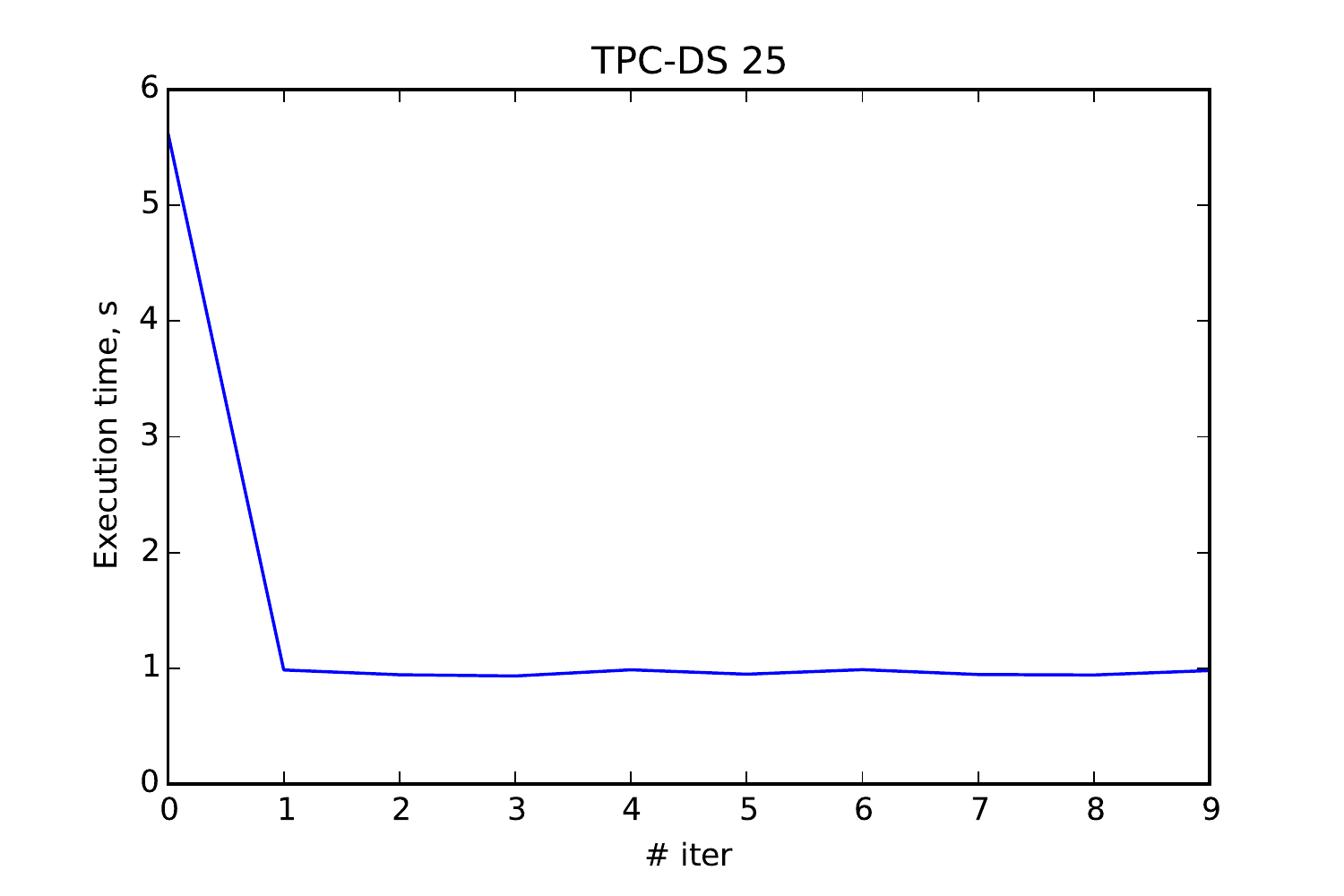}\hfill%\
\includegraphics[width=0.3\linewidth]{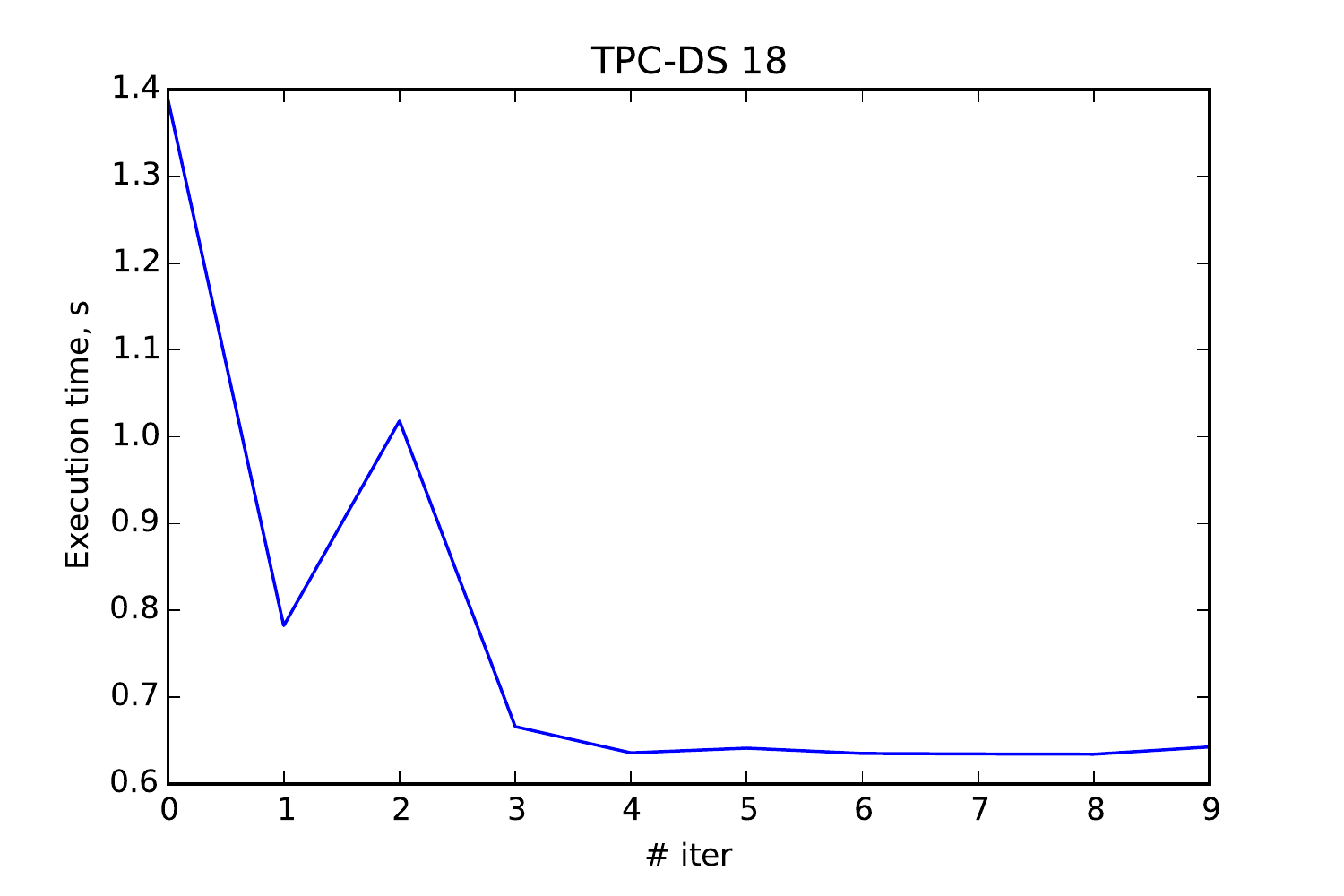}\hfill%\
\includegraphics[width=0.3\linewidth]{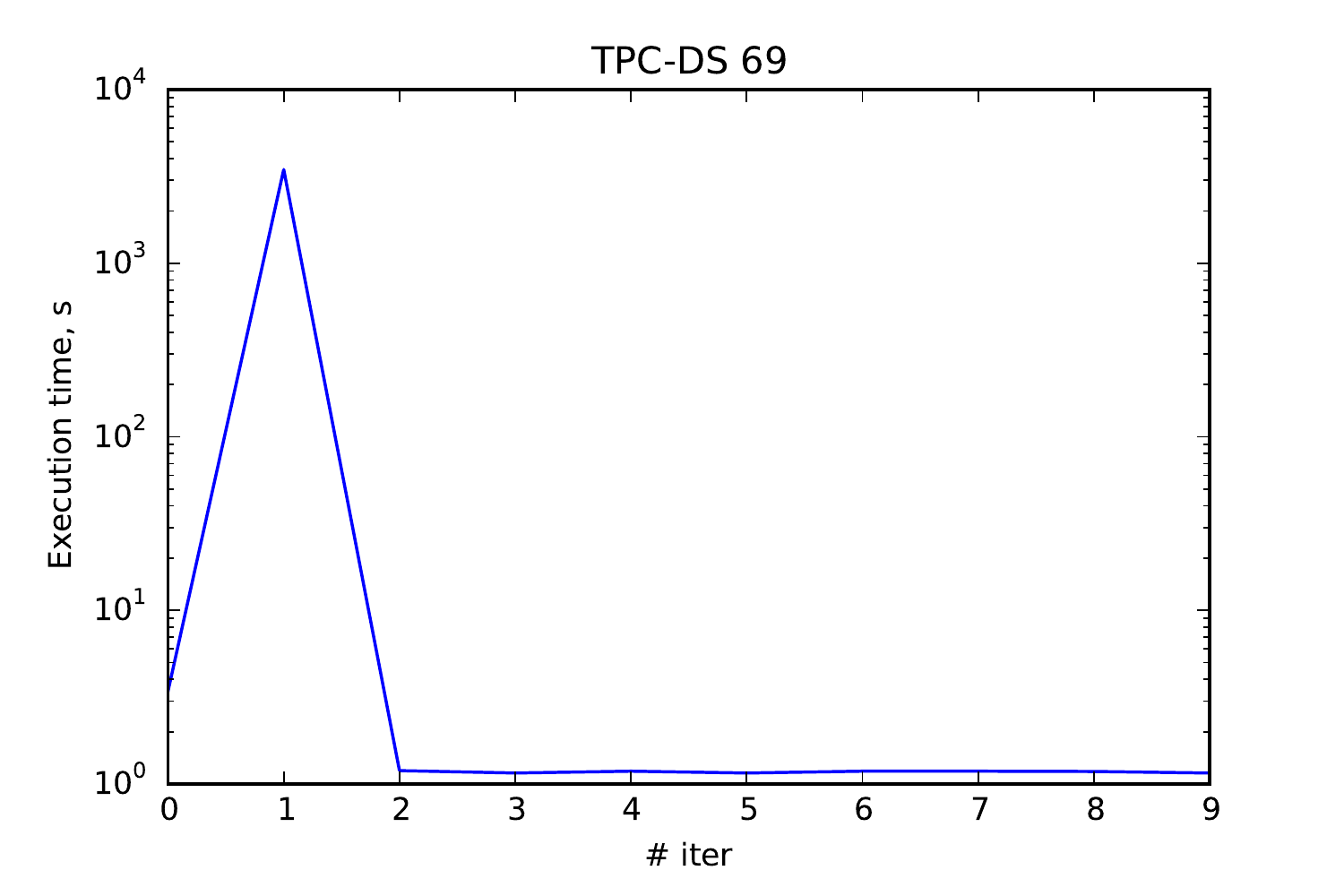}\\
\caption{Query execution time during learning procedure}
\label{fig:et_dynamics}
\end{figure*}

The only theoretical guarantee we have is that our method will converge in finite number of steps under some reasonable constraints. Also we know that performance does not decrease after the convergence with good enough cost model. Nevertheless, we have no such guarantees during learning procedure. On figure~\ref{fig:et_dynamics} we can see that query execution time may decrease monotonically, decrease not-monotonically or even grow much more than initial execution time before the final decrease. The most common case of learning for TPC--DS is the convergence in one step which means monotonically not-increase of execution time, but other variants are also possible.

\begin{figure}
\centering
\includegraphics[width=0.67\linewidth]{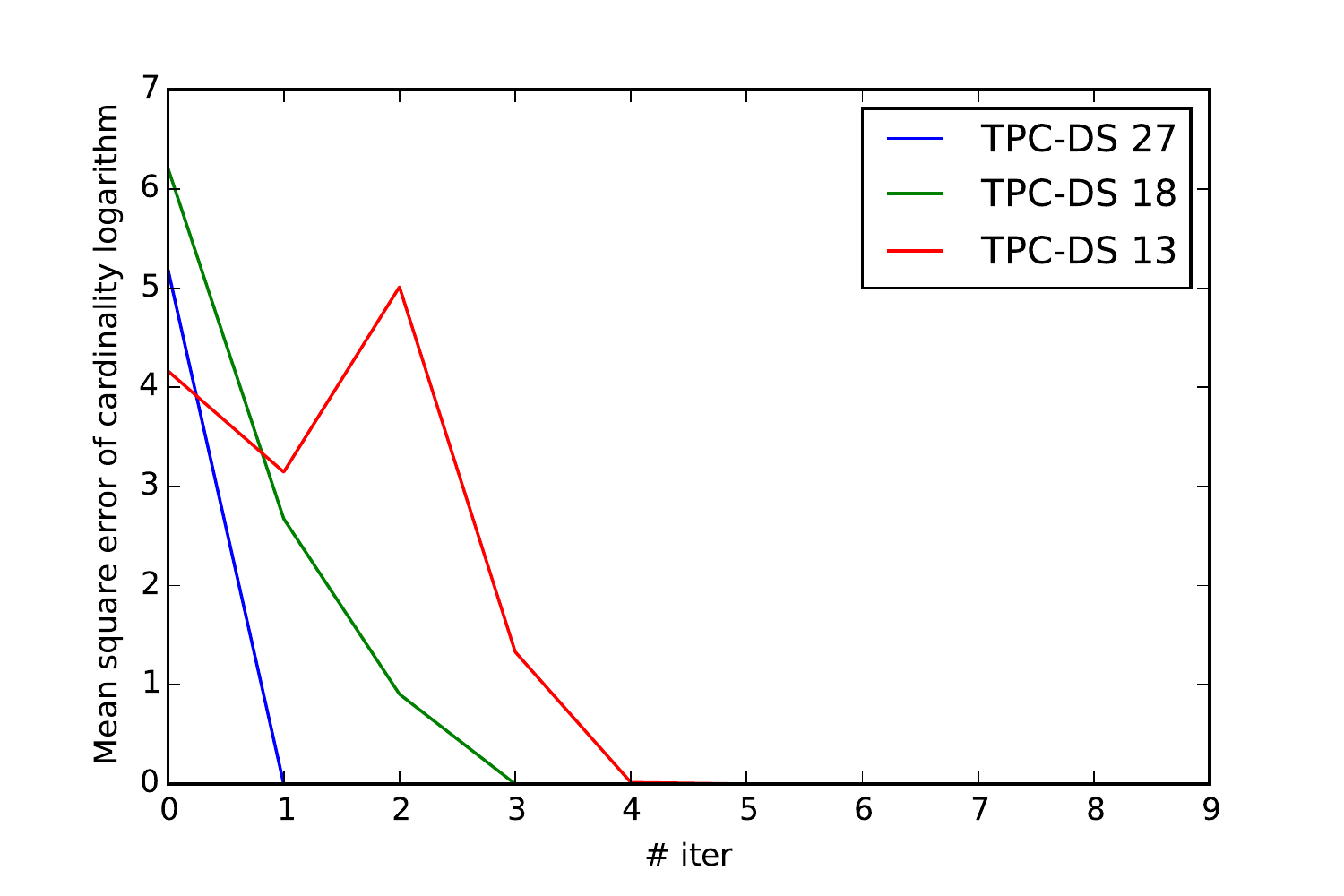}\\
\caption{Cardinality prediction quality during learning procedure}
\label{fig:cardinality_dynamics}
\end{figure}

The cardinality prediction quality also not necessarily grows monotonically, but after convergence this quality is not worse than before learning (see figure~\ref{fig:cardinality_dynamics}).

We can see that our method converges in a few steps or a few dozens of steps in practice.

The phenomenon of possible bad cardinality predictions and therefore possible bad performance during learning procedure is the principle drawback of methods which try to learn joint data distribution based on query execution statistics only. Nevertheless there are ways to avoid this drawback. Query re-optimization is the most straight-forward way to do this. It proposes not to execute bad plans completely, but to use statistics from partially-executed bad plans to improve cardinality prediction and then restart query execution with a better plan. The plan space exploration allows to pre-compute some cardinalities in background which can improve cardinality for query execution.

\section{Future work}\label{sec:future}

The proposed method can significantly increase DBMS performance, but it also has drawbacks and rooms for improvement waiting for further research.

\paragraph*{Feature space construction} In this paper we don't investigate all possible reductions of cardinality estimation problem to machine learning problem. Probably there is a better way to construct feature space for machine learning problem which leads to better DBMS performance. For example, one can explore the reduction which consider clause structure in a better way.

\paragraph*{Dynamic data and workload} The proposed method is designed for nearly static database and queries. If database and queries are changed significantly, it would be better for administrator to restart the method. The problem for the research is to adapt the proposed method for dynamically changing database and workload, which includes determining, removing and updating outdated statistics. The proposed fixed-memory $k$ nearest neighbours method performs such adaptation, but firstly this quality of method was not tested yet, and secondly we believe that method with faster adaptation rate may be developed.

\paragraph*{Query re-optimization} The proposed in this paper technique allows to use query execution statistics for the better plan choice. Nevertheless, sometimes (during learning process or when data in database is changed) query optimizer mistakes too much. That leads to bad plans. Often we have no need to execute these plans completely to understand that there was a mistake in query optimization. That is why we can interrupt and rollback the execution of such plans, use the obtained execution statistics to choose better plans, and then execute these better plans. This induces two problems for further research: how to use statistics from the interrupted query and how to automatically determine whether the query optimizer chose a bad plan and when to interrupt it. Potentially this research leads to faster learning of the predictive model and saves us from the full execution of really bad plans.

We propose the idea how to implement this principle. Because the common case of bad plan is caused by underestimation of the cardinality, we propose to save to statistics only those nodes from the interrupted query in which cardinality estimation is much less than the real number of tuples. Also it is better to proceed the information from the interrupted queries separately, for example forgetting them faster, because their information is not precise. We can consider that the plan is bad if the cardinality estimation in any its node is less by ten times than the real number of tuples in there. Nevertheless, this idea needs more detailed elaboration and the experimental evaluation.

\paragraph*{Plans space exploration} Sometimes the combination of standard uncertain predictions and machine learning predictions leads optimization method to choose not-optimal plan. That may be fixed with the use of different exploration techniques. The point is that sometimes we can execute not-optimal plans to obtain the information about the real cardinalities of some sets of clauses (maybe it is better to do in background). This technique allows us to guarantee the fastest plan choice if the cost model is good enough.

\section{Conclusions}

There are three main points in this paper.
Firstly, we show that the plans constructed by the standard query optimizer based on clauses independence assumption are far from optimal. One can observe that the more complicated query structure is, the more speed up for this query is potentially available with the use of better query execution plan.
Secondly, we not only state that the better plans are available, but also propose the methods for finding them, evaluate these methods on different benchmarks, perform the basic theoretical analysis of these methods. The main idea of the proposed method is to improve the node cardinality estimation.
Thirdly, the novelty of our approach lies in using for the cardinality estimation not only pre-computed statistics, but the execution statistics of previous queries also.
We cannot state that the proposed methods completely solve the query optimization problem.
Nevertheless we show that they can significantly increase the quality of this problem solution and therefore significantly increase the DBMS performance.

% ensure same length columns on last page (might need two sub-sequent latex runs)
\balance

\section{Acknowledgments}
This work was supported by Postgres Professional. Authors thank Oleg Bartunov for the idea proposal and for the talks. Authors thank Boris Novikov for the commentaries and advice on the paper text.

\bibliographystyle{abbrv}
\bibliography{refs}
% ****************** APPENDIX **************************************
% Example of an appendix; typically would start on a new page
%\pagebreak

\begin{appendix}

\section{What is worse: cardinality or cost estimation?}\label{sec:cardinality_or_cost}

\begin{figure*}
\centering
\includegraphics[width=0.49\linewidth]{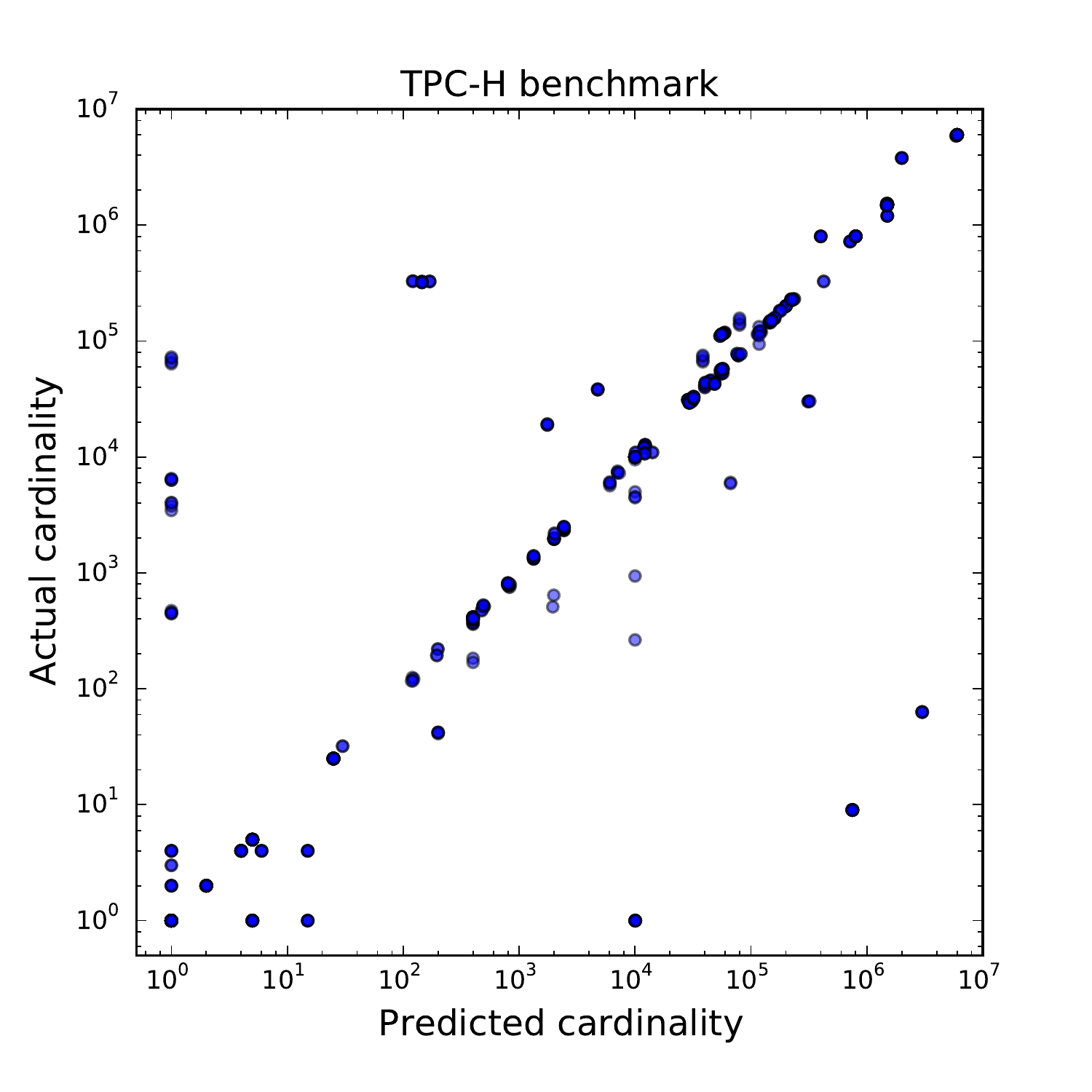}\hfill%\
\includegraphics[width=0.49\linewidth]{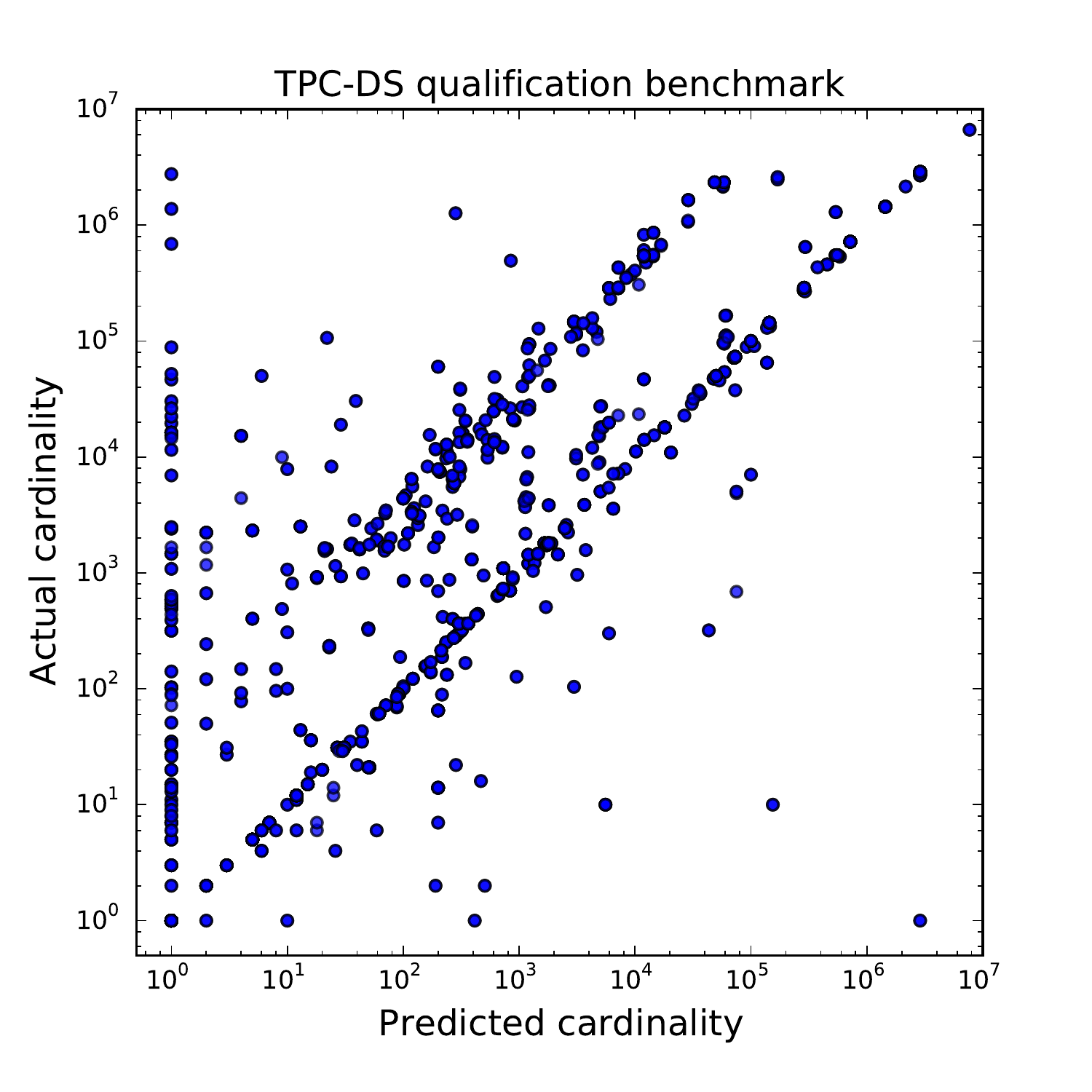}\\
\caption{Quality of cardinality prediction using standard PostgreSQL estimator}
\label{fig:tpc_cardinality}
\end{figure*}

To answer this question we use TPC--H benchmark and TPC--DS qualification benchmark with scale factor 1Gb both. Cardinality estimation results are available on figure~\ref{fig:tpc_cardinality}. Each scan or join plan node is represented as a point on the figure.

\begin{figure*}
\centering
\includegraphics[width=0.49\linewidth]{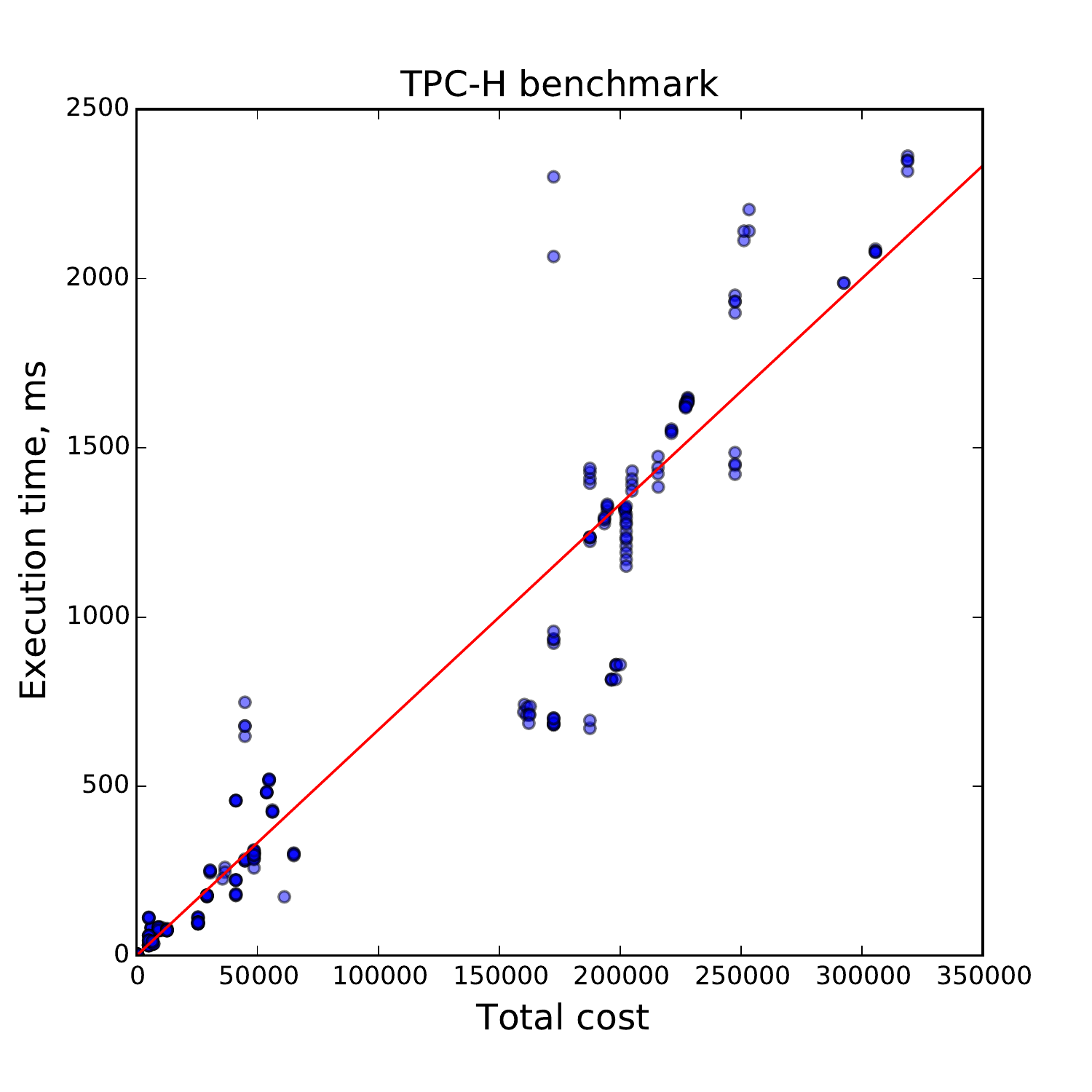}\hfill%\
\includegraphics[width=0.49\linewidth]{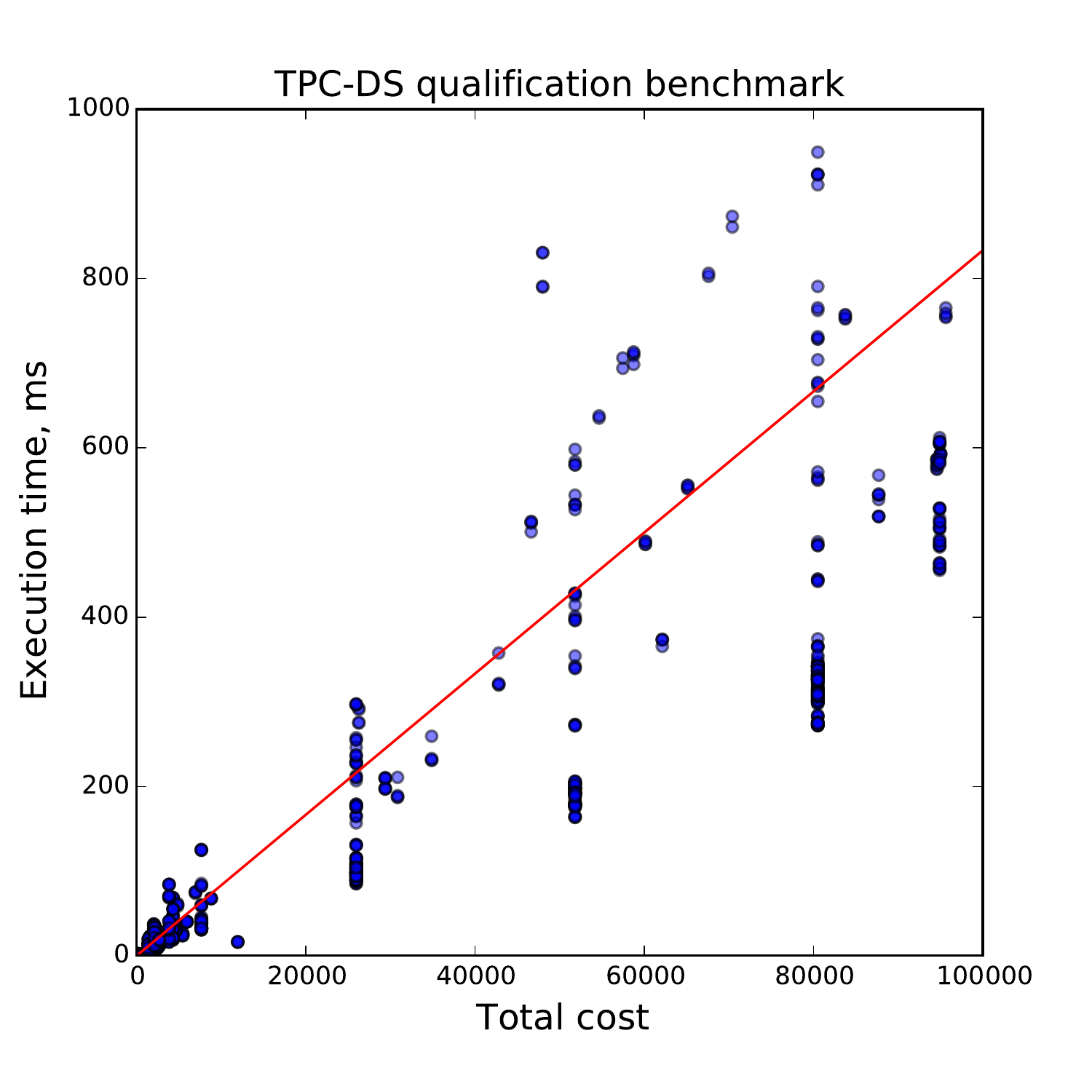}\\
\caption{Quality of PostgreSQL cost model}
\label{fig:tpc_et}
\end{figure*}

For evaluation of PostgreSQL cost model we selected only those nodes in which for them and for their children the cardinality estimation is correct with maximal relative error 10\%. Each such scan or join node is represented as a point on the figure~\ref{fig:tpc_et}. We approximate the linear dependence between cost and execution time using red line on the figure.

On figure~\ref{fig:tpc_cardinality} we see several order errors. On figure~\ref{fig:tpc_et} there are only several times errors. That induces following conclusion: the cost model is more or less adequate with the correct cardinality prediction. The cardinality prediction causes the most significant errors in query optimization.

%\section{Tuning parameters for machine learning methods}\label{sec:ml_tuning}

\section{Fixed-memory nearest neighbour \\ nonparametric regression}\label{sec:OkNN}

$k$ nearest neighbours is one of the most well-known algorithm in machine learning. Its regression modification works using the following formula:
\begin{equation}\label{eq:knn}
    \hat y = \frac{\sum\limits_{i=1}^{k}y_{(i)} \cdot sim(x_{(i)}, x)}{\sum\limits_{i=1}^{k}sim(x_{(i)}, x)}
\end{equation}
In this formula $x$ is the feature vector of new object and $\hat y$ is our prediction for it. $k$ is the number of the nearest neighbours to consider. $sim$ is the similarity function in the feature space. $x_{(i)}$ and $y_{(i)}$ are objects of tthe raining set sorted by decrease of their similarity with the new object $x$.

One can learn more about $k$ nearest neighbours algorithm in~\cite{bishop2006patternkNN}.
The benefit of this method is a lack of strong assumptions about the shape of the true regression function.
Nevertheless, we cannot use this method straightforward.

%The only paper for online learning $k$ nearest neighbours classifier is~\cite{bermejo1999line}.

We consider the following learning procedure: objects and their true target values are becoming available sequentially. After appending of each object the model is updated.

The limitation of learning procedure is that we cannot store all objects, because there are too many of them. Another reason to store less objects is because it decreases the computational complexity of finding the nearest ones.

Let us define the maximal available for the storage number of objects as $K$.

Before the limit of $K$ objects is reached we act almost like in the original $k$ nearest neighbour regression and add the new objects to the training set. The only difference is that during the learning procedure we want to minimize the amount of objects to store. So we don't want to store similar objects. We believe in data locality, i. e. for similar objects their target values are also similar. That leads to the following heuristics for adding a new object into the training set: if in the set there is an object $x_i$ in radius $\delta$ from the new object $x$ then we just modify $x_i$ and $y_i$.
\begin{gather}
x_i \leftarrow x_i + \eta (x - x_i) \\
y_i \leftarrow y_i + \eta (y - y_i)
\end{gather}
We call this heuristic object filtering. $\eta$ is the stochastic gradient learning rate.

Another way to minimize the number of stored objects is not to store objects for which we already have nearly perfect prediction using previous objects. We don't use this way in our implementation.

After limit $K$ is reached we cannot store new objects. Nevertheless, we have to fit our regressor with the new data. We have to modify stored objects and their target values somehow for this fitting. Stochastic gradient descent may be used to do that. It works for an arbitrary differentiable loss function $l(\hat y, y)$ and similarity function $sim(x_{(i)}, x)$. The point of method is taking the derivatives of $l(\hat y, y)$ with respect to $x_{(i)}$ and $y_i{(i)}$ using $\hat y$ from formula~\ref{eq:knn}.

\begin{gather}
\frac{\partial l(\hat y, y)}{\partial y_{(l)}} = \frac{\partial l(\hat y, y)}{\partial \hat y} \cdot \frac {sim(x_{(l)}, x)}{\sum\limits_{i=1}^{k} sim(x_{(i)}, x)} \\
\frac{\partial l(\hat y, y)}{\partial x_{(l)}} = \frac{\partial l(\hat y, y)}{\partial \hat y} \cdot \frac{\partial sim(x_{(l)}, x)}{\partial x_{(l)}} \cdot \frac{y_{(l)} - \hat y}{\sum\limits_{i=1}^{k} sim(x_{(i)}, x)}
\end{gather}

We perform a step contrariwise this gradient using the learning rate $\eta$. That is how one can use stochastic gradient descent for $k$ nearest neighbour regression.

Note that after limit $K$ is reached we store the virtual objects instead of the real ones. That means that stored objects $x_i$ not necessarily corresponds to any of objects obtained during the learning procedure. The same statement holds true for the target values. Nevertheless, regressor built on such virtual objects stores information about all obtained real objects in such way that minimizes loss function for the predictions for the new objects.

In our implementation of the algorithm we use the following formulas for loss and similarity
\begin{gather}
    l(\hat y, y) = \frac12(\hat y - y)^2 \\
    sim(x_i, x) = \frac{1}{0.1 + {||x_i - x||}_{2}}
\end{gather}

The obtained method fulfills memory limitation and doesn't increase update time significantly. On figure~\ref{fig:compare_01} we can see that it works significantly better than just $k$ nearest neighbours with the last $K$ objects stored and object selection heuristics.

%\section{Theoretical properties}\label{sec:theorems}

%// TBD

\end{appendix}

\end{document}